\documentclass{article}

\usepackage[margin=1.25in]{geometry}

\usepackage{graphicx} % Required for inserting images

\usepackage{amsmath,amssymb,amsthm}
\usepackage[labelfont=bf]{caption}

\usepackage{enumitem}
\usepackage{soul}

\usepackage{helvet}
\usepackage{sectsty}
\allsectionsfont{\sffamily}
\usepackage{newtxtext,newtxmath}

\usepackage[dvipsnames]{xcolor}

\definecolor{c0}{HTML}{1d1d1d}
\definecolor{c1}{HTML}{173359}
\definecolor{c2}{HTML}{8f5a39}
\definecolor{c3}{HTML}{a8332b}
\definecolor{c4}{HTML}{4c825c}

\usepackage{algorithm, algpseudocode, algorithmicx}

\algrenewcommand{\algorithmiccomment}[1]{\hfill \textcolor{gray}{$\vartriangleright$ \textit{#1}}}
\usepackage{etoolbox}
\algrenewcommand\alglinenumber[1]{\footnotesize #1:}
\makeatletter
\expandafter\patchcmd\csname\string\algorithmic\endcsname{\itemsep\z@}{\itemsep=5pt}{}{}
\makeatother

\usepackage{listings}
\usepackage{sectsty}

\sectionfont{\bfseries\sffamily\large\selectfont}
\subsectionfont{\bfseries\sffamily\normalsize\selectfont}
\subsubsectionfont{\bfseries\sffamily\small\selectfont}
\paragraphfont{\bfseries}

\usepackage{titlesec}
\titlespacing*{\section}{0pt}{.5em}{0em}
\titlespacing*{\subsection}{0pt}{.5em}{0em}
\titlespacing*{\subsubsection}{0pt}{.5em}{0em}

\usepackage{float}

\usepackage{booktabs}

\usepackage{hyperref}
\hypersetup{
    colorlinks,
    linkcolor=c2,
    citecolor=c2,
    urlcolor=c2
}
\usepackage[nameinlink,capitalize,noabbrev]{cleveref}
\crefformat{equation}{#2(#1)#3}
\crefmultiformat{equation}{{}#2(#1)#3}{ and~#2(#1)#3}{, #2(#1)#3}{ and~#2(#1)#3}

\newtheorem{theorem}{Theorem}

\newtheorem{lemma}[theorem]{Lemma}
\newtheorem{proposition}[theorem]{Proposition}

\theoremstyle{definition}
\newtheorem{definition}[theorem]{Definition}
\newtheorem{remark}[theorem]{Remark}

\numberwithin{theorem}{section}
\numberwithin{equation}{section}

\renewcommand{\vec}{\mathbf}

\newcommand{\cond}{\operatorname{cond}}
\newcommand{\T}{\mathsf{T}}

\newcommand{\smax}{\sigma_{\textup{max}}}
\newcommand{\smin}{\sigma_{\textup{min}}}

\newcommand{\argmin}{\operatornamewithlimits{argmin}}

\newcommand{\nnz}{\operatorname{nnz}}

\usepackage{tikz}

\usepackage[style=alphabetic,url=false]{biblatex}
\begin{document}
\color{c0}

\title{\Large\sffamily{\textbf{GPU-Parallelizable Randomized Sketch-and-Precondition for Linear Regression using Sparse Sign Sketches}}}

%alphabetical
\author{Tyler Chen\thanks{Email: \url{tyler.chen@jpmchase.com}}\and
Pradeep Niroula\and
Archan Ray\and
Pragna Subrahmanya \and
Marco Pistoia\and
Niraj Kumar\thanks{Principal Investigator. Email: \url{niraj.x7.kumar@jpmchase.com}}}
\date{Global Technology Applied Research, JPMorgan Chase, New York, NY 10001, USA}

\maketitle

\begin{abstract}
A litany of theoretical and numerical results have established the sketch-and-precondition paradigm as a powerful approach to solving large linear regression problems in standard computing environments.
Perhaps surprisingly, much less work has been done on understanding how sketch-and-precondition performs on graphics processing unit (GPU) systems. 
We address this gap by benchmarking an implementation of sketch-and-precondition based on sparse sign-sketches on single and multi-GPU systems.
In doing so, we describe a novel, easily parallelized, rejection-sampling based method for generating sparse sign sketches.
Our approach, which is particularly well-suited for GPUs, is easily adapted to a variety of computing environments.
Taken as a whole, our numerical experiments indicate that sketch-and-precondition with sparse sign sketches is particularly well-suited for GPUs, and may be suitable for use in black-box least-squares solvers.
\end{abstract}

\section{Introduction}

We are interested in solving the large over-determined linear least squares problem
\begin{equation}
    \min_{\vec{x}\in\mathbb{R}^{n}} \|\vec{b} - \vec{A}\vec{x}\|
    ,\quad \vec{A}\in\mathbb{R}^{m\times n}, \quad\vec{b}\in\mathbb{R}^m
    ,\quad m\gg n\gg1,
    \label{eqn:lstsq}
\end{equation}
one of the core tasks in numerical linear algebra.
% Such problems arise naturally in a broad range of applications including approximation \cite{}, tomography  \cite{huang_wang_lee_chen_12,lee_haung_dennis_chen_wang_13,vanaarle_palenstijn_debeenhouwer_altantzis_bals_batenburg_sijbers_15}, and as a sub-routine within Krylov subspace methods (KSMs) for large sparse linear systems
% \cite[etc.]{balabanov_grigori_22,nakatsukasa_tropp_24,landman_brown_chung_nagy_25}.
The classical approach to solving such problems is by direct factorization methods such as Householder QR.
However, direct methods, which require $O(mn^2)$ arithmetic operations, are intractable when $m$ and $n$ are large. 
In such situations, Krylov subspace methods (KSMs), such as LSQR, are a commonly used to obtain an approximate solution to \cref{eqn:lstsq}.
The total cost of such algorithms is determined by the \emph{cost per iteration} and the \emph{total number of iterations}. 
The per-iteration cost of KSMs is dominated by a matrix-vector product with each $\vec{A}$ and $\vec{A}^\T$, which amounts to $O(\nnz(\vec{A})) \leq O(mn)$ easily parallelizable arithmetic operations.
Therefore, KSMs are efficient so long as the total number of iterations is reasonably small.
Unfortunately, the total number of iterations required by KSMs can be very large if $\vec{A}$ is poorly conditioned, presenting a practical barrier in many situations.

To try to reduce the required number of iterations, it is common to use a technique called \emph{preconditioning}.
The aim of preconditioning is to transform \cref{eqn:lstsq} into an equivalent problem on which iterative methods will converge more rapidly \cite[etc.]{saad_03}.
Specifically, given an invertible matrix $\vec{M}\in\mathbb{R}^{n\times n}$ (called a \emph{preconditioner}), one can construct the (right) preconditioned regression problem 
\begin{equation}
    \min_{\vec{x}\in\mathbb{R}^{n}} \|\vec{b} - (\vec{A}\vec{M})\vec{y}\|
    ,\qquad
    \vec{x} = \vec{M}\vec{y}.
    \label{eqn:lstsq_prec}
\end{equation}
One easily verifies that \cref{eqn:lstsq_prec} has the same solution as \cref{eqn:lstsq}.
Notably, however, if $\vec{A}\vec{M}$ is well-conditioned, then KSMs will converge rapidly on \cref{eqn:lstsq_prec}.
Moreover, $\vec{A}\vec{M}$ does not need to be formed explicitly. 
Instead, matrix-vector products with $\vec{A}\vec{M}$ and $(\vec{A}\vec{M})^\T$ are respectively performed by products with $\vec{M}$ and then $\vec{A}$ or with $\vec{A}^\T$ and then $\vec{M}^\T$, which does not incur much additional cost per iteration.
As such, the efficacy of preconditioning depends primarily on balancing the improvement in the number of iterations with the cost of constructing and applying $\vec{M}$.

Perhaps the most promising general approach to constructing a preconditioner for \cref{eqn:lstsq} is based the \emph{sketch-and-precondition} paradigm \cite{rokhlin_tygert_08}.
To build the preconditioner, a sketching matrix $\vec{S}\in\mathbb{R}^{d\times m}$ is sampled randomly (from an appropriate distribution) and applied to $\vec{A}$.
The sketch $\vec{S}\vec{A}$ is processed in order to build a preconditioner $\vec{M}$.
% so that
% \begin{enumerate}[noitemsep,topsep=-.5em,label=(\roman*)]
% \item $\vec{S}$ can be efficiently applied to $\vec{A}$, 
% \item  $\vec{S}\vec{A}\in\mathbb{R}^{d\times n}$ is much smaller than $\vec{A}$ ($d\ll m$), and 
% \item $\vec{A}^\T\vec{A} \approx (\vec{S}\vec{A})^\T (\vec{S}\vec{A})$.
% \end{enumerate}
% The sketched matrix $\vec{S}\vec{A}$ can then be efficiently processed using classical direct methods in order to obtain a preconditioner $\vec{M}$.
% In particular, the Cholesky factorization of $(\vec{S}\vec{A})^\T (\vec{S}\vec{A})$ (obtained via a QR fatorization of $\vec{S}\vec{A}$) yields a preconditioner.
As we discuss further in \cref{sec:bacground}, randomized preconditioning methods for \cref{eqn:lstsq} enjoy strong theoretical runtime guarantees, typically requiring $\tilde{O}(mn + n^3)$ operations.\footnote{We use $\tilde{O}$ to hide poly-logarithmic factors in $m$, $n$, and the target accuracy.}
The benefits of such methods extends to practical computations as well.
For instance, the \emph{Blendenpik} \cite{avron_maymounkov_toledo_10}, % and \emph{LSRN} \cite{meng_saunders_mahoney_14} solvers
an early implementation of this framework, was demonstrated to outperform LAPACK's classical solvers, even on moderately sized problems.
Such preconditioning techniques are also at the core of large-scale numerical software development, including the next-generation \emph{RandLAPACK} \cite{muarray_etal_23}.

\begin{figure}[ht]
    \centering
    \includegraphics[scale=.7]{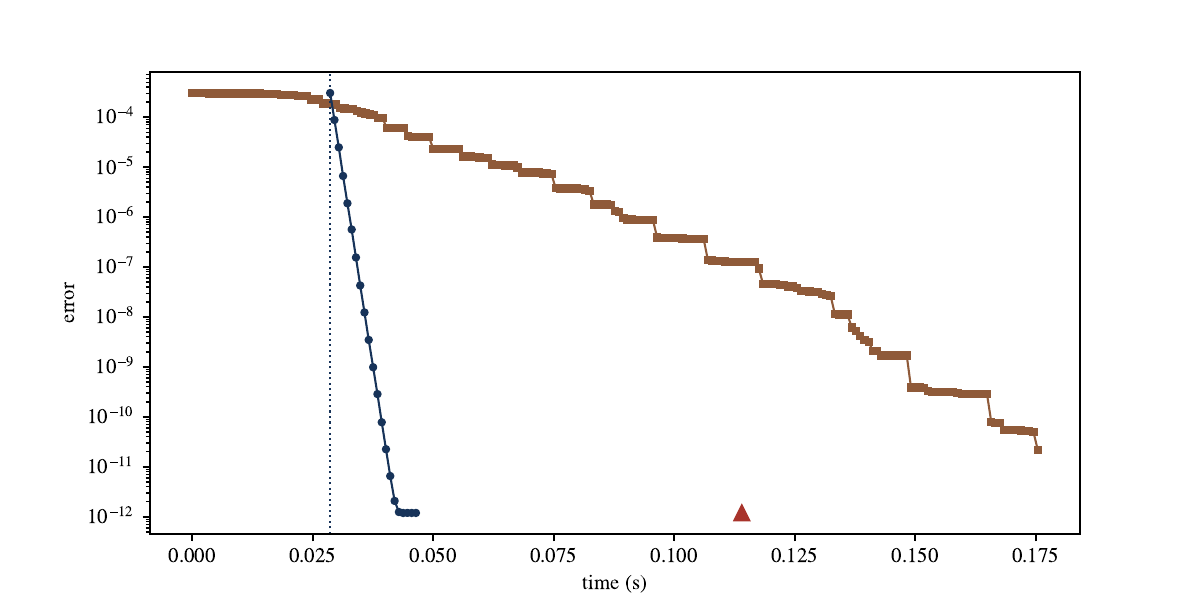}
    \caption{Comparison of accuracy/runtime for a direct solver ({\protect\includegraphics[scale=1]{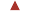}}), iterative method without a preconditioner ({\protect\raisebox{.25pt}{\protect\includegraphics[scale=0.85]{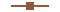}}}), and sketch-and-precondition ({\protect\raisebox{.25pt}{\protect\includegraphics[scale=0.85]{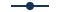}}}), on a dense, moderately sized, moderately conditioned least squares problem ($m=10^5$, $n=6\times 10^2$, $\cond(\vec{A}) = 10^3$).
    While sketch-and-precondition requires some time to construct the preconditioner (indicated by the vertical dotted line), this is more than made up for by the improvement to convergence provided by the preconditioner.
    }
    \label{fig:intro}
\end{figure}

In \cref{fig:intro} we provide a simple example illustrating the potential of the sketch-and-preconditioner paradigm on the GPU.
Even on a dense, moderately sized, moderately conditioned least squares problem ($m=10^5$, $n=10^2$, $\cond(\vec{A}) = 10^3$), sketch-and-precondition exhibits substantial speedups over direct methods and unpreconditioned KSMs.
On  larger, more poorly conditioned, or sparse problems, the advantages of sketch-and-precondition over the other methods is even more pronounced.

\subsection{GPU and multi-GPU computations}

Graphics processing units (GPUs) are designed to be able to perform many simple tasks in parallel. 
As such, they are extremely well suited for performing basic matrix operations, and their use has become widespread in the computational sciences.
Each GPU has its own memory that it can access quickly. 
However, this memory is limited, and moving data from the system memory to the GPU's memory is expensive relative to on-device arithmetic operations.
Many computers are equipped with multiple GPUs that can be used together to accelerate a single application.
However, doubling the number of GPUs available does not necessarily halve the runtime.
In particular, the benefits of additional compute resources must be balanced against the increased communication overhead \cite{owens_houston_luebke_green_stone_phillips_08,dongarra_tomov_luszczek_kurzak_gates_yamazaki_anzt_haidar_abdelfattah_17}.

\subsection{Related work}

% \cite{ozaslan_pilanci_arikan_19}

% For historical reasons, variants of \cref{alg:main} that use a gradient method as the iterative method are often referred to \emph{iterative sketching} methods \cite{?,epperly_24,epperly_meier_nakatsukasa_24}.
% In \cref{eqn:iterative_sketching} we provide a more detailed discussion.

The sketch-and-precondition paradigm was introduced in \cite{rokhlin_tygert_08}, and early software implementations demonstrated that the method was competitive with highly optimized classical solvers in many settings \cite{avron_maymounkov_toledo_10,meng_saunders_mahoney_14}.
% Sketch-and-precondition has also been used in many theoretical advancements of algorithms for \cref{eqn:lstsq}, including the first optimal least squares algorithms \cite{clarkson_woodruff_13,chenakkod_derezinski_dong_rudelson_24}.
Recent work has studied the numerical stability of sketch-and-precondition methods and shown that proper implementation results in algorithms that, under reasonable assumptions, are just as stable as classical direct methods (i.e., forward and backward stable) \cite{meier_nakatsukasa_townsend_webb_24,epperly_24,epperly_meier_nakatsukasa_24}.
In total, it is reasonable to expect that sketch-and-precondition methods will soon be able to serve as drop-in replacement for classical least-squares methods.

While most work on sketch-and-precondition has been limited to standard computing devices (e.g. a single node), a number of papers have considered sketch-and-precondition on distributed memory clusters \cite{meng_saunders_mahoney_14,iyer_avron_kollias_ineichen_carothers_drineas_15,iyer_avron_kollias_ineichen_carothers_drineas_19}.
While such systems share many features with multi-GPU systems, much less work has been done on sketch-and-precondition on GPU-enabled systems. 
In fact, the only work we are aware of that considers sketch-and-precondition on GPUs is \cite{georgiou_boutsikas_drineas_anzt_23}, which studies the use of Gaussian sketching on a single GPU, accelerated by using reduced-precision to construct and apply the sketching matrix.
Dense sketching matrices do not scale well (in theory or practice), and we would caution against their use in most settings. 

Separately, several parallel or GPU variants of the LSQR algorithm have been developed \cite{huang_wang_lee_chen_12,lee_haung_dennis_chen_wang_13,cesare_etal_22}.
These variants could be used in an off-the-shelf manner within sketch-and-precondition methods.
However, due to the relatively higher costs of the matrix-vector product each iteration, standard variants of Krylov subspace methods for very tall linear least squares problems are substantially easier to parallelize than those for sparse linear systems
\cite{li_saad_12,heroux_dongarra_13,carson_15}, so the benefits of more complicated variants are not as clear as for sparse linear systems.

Finally, we note the work on other randomized sketching-based algorithms such as the randomized SVD and preconditioned Cholesky QR on (multi)-GPU systems \cite{meng_saunders_mahoney_14,struski_morkisz_spurek_bernabeu_rodriguez_trzcinski_24,baboulin_donfack_kaya_mary_robeyns_24,higgins_szyld_boma_yamazaki_24}.
As with sketch-and-precondition, generating and applying a sketching matrix is one of the dominant costs in these algorithms.
Randomized linear algebra algorithms are often used with the explicit goal of increasing data-parallelism, possibly even at the cost of increasing the number of arithmetic operations \cite[etc.]{meng_saunders_mahoney_14,fan_guo_lin_21,balabanov_22,melnichenko_balabanov_murray_demmel_mahoney_luszczek_24,higgins_szyld_boma_yamazaki_24}.
Many of our findings about sparse sign sketches may be applicable to these settings as well.

% Finally, we note that KSMs are also used to solve (large) sparse linear systems, and there has been substantial work to understanding how to effectively scale such algorithms on distributed computers \cite{?}.
% In such settings, vector operations typically have a similar arithmetic cost to matrix-vector products but much higher communication costs. 

\subsection{Notation}

We denote the smallest and largest eigenvalues of $\vec{A}$ by $\smin(\vec{A})$ and $\smax(\vec{A})$ respectively, and the condition number of $\vec{A}$ is $\cond(\vec{A}) = \smax(\vec{A})/\smin(\vec{A})$.
The total number of nonzero entries of $\vec{A}$ is $\nnz(\vec{A})$.
Throughout, $\|\cdot\|$ indicates the Euclidian norm for vectors and spectral norm for matrices.
We write $[d]$ to indicate the set $\{1,\ldots, d\}$. 

\section{Sketch-and-precondition overview}
\label{sec:bacground}

A meta-algorithm for the sketch-and-precondition framework for solving \cref{eqn:lstsq} is described in \cref{alg:main}. 
We emphasize that sketch-and-precondition is a very general framework and many different implementations are possible.
The two main design choices are which \emph{sketching distribution} and which \emph{iterative method} to use, and different choices result in different implementations. 
\begin{itemize}
    \item \textbf{Sketching distribution}~
Common choices include
Gaussian sketches \cite{indyk_98,halko_martinsson_tropp_11,woodruff_15}, subsampled fast trigonometric transform sketches \cite{woolfe_libety_rokhlin_tygert_08,halko_martinsson_tropp_11,rokhlin_tygert_08,avron_maymounkov_toledo_10, tropp_11, boutsidis_gittens_13}, and increasingly, sparse sign sketches\footnote{Sparse sign sketches are also  referred to as short-axis-sparse sketching operators (SASO) \cite[\S2.4.1]{muarray_etal_23} and are a generalization of count sketch \cite{woodruff_15}.} \cite{meng_mahoney_13,clarkson_woodruff_13,nelson_nguyen_14,woodruff_15,tropp_yurtsever_udell_cevher_19}.
For additional overviews, we turn readers to \cite[\S2]{muarray_etal_23} and \cite[\S8,9]{martinsson_tropp_20}.

\item \textbf{Iterative method}~
By far the most common choice of iterative methods are Krylov subspace methods (KSMs) which include commonly used  algorithms like conjugate gradient (CG) and LSQR \cite{paige_saunders_82,saad_03} as well as simpler methods like gradient descent and heavy ball momentum.\footnote{For historical reasons, variants of \cref{alg:main} that use a gradient method as the iterative method are often referred to \emph{iterative sketching} methods.
We provide a more detailed discussion on this connection in \cref{eqn:iterative_sketching}.}
Stochastic iterative methods can also be used \cite{yang_chow_re_mahoney_18,chenakkod_derezinski_dong_rudelson_24}.
\end{itemize}

\begin{algorithm}[ht!]
\caption{Sketch-and-precondition}
\label{alg:main}
	\begin{algorithmic}[1]
		\Require Matrix $\vec{A}\in\mathbb{R}^{m\times n}$, vector $\vec{b}\in\mathbb{R}^m$, target accuracy $\varepsilon$
        \State Sample sketching matrix $\vec{S}$ from $\Call{SketchingDistribution}{}$
        \Comment{Generate sketching matrix}
        \State Form $\vec{Y} = \vec{S}\vec{A}$ 
        \Comment{Apply sketching matrix}
        \State Factor $\vec{Q},\vec{R} = \Call{QR}{\vec{Y}}$ 
        \Comment{Factor sketch}
        \State $\vec{M} = \vec{R}^{-1}$
        \Comment{Get preconditioner}
        \label{alg:main:qr}
        \State $\vec{x}_0 = \vec{M}\vec{Q}^\T\vec{S}\vec{b}$ 
        \Comment{Get initial guess}
        \State $\widehat{\vec{x}} = \Call{IterativeMethod}{\vec{A},\vec{M},\vec{b},\vec{x}_0,\varepsilon}$
        \Comment{Run preconditioned iterative method}
		\Ensure Approximate solution $\widehat{\vec{x}}$ to \cref{eqn:lstsq}
	\end{algorithmic}
\end{algorithm}

In this paper, we focus on sparse sign sketches and LSQR, which we believe are respectively the best choices of sketching distribution and iterative method in most situations, and particularly when parallel resources are available.
Further discussion and a numerical comparison against other sketching distributions and iterative methods are given in \cref{sec:which_sketch} and \cref{sec:KSMs} respectively.

A typical theoretical guarantee for sketch-and-precondition with sparse sign sketches and LSQR looks like the following:\footnote{This bound is included to give a general sense of the type of theoretical guarantees available. More precise theory is known and can be found in the references throughout this section.}
\begin{theorem}\label{thm:main}
Consider the least squares problem \cref{eqn:lstsq} with solution $\vec{x}_\star$.
Implement \cref{alg:main} with sparse sign sketches and LSQR.
Then, using 
\begin{equation*}
    \tilde{O}\big((\nnz(\vec{A}) + n^2)\log(1/\varepsilon) + n^3\big)
\text{~operations,}
\end{equation*}
the output $\widehat{\vec{x}}$ satisfies, with probability exceeding $99/100$
\begin{equation*}
\| \vec{A}(\vec{x}_\star - \widehat{\vec{x}}) \|
\leq \varepsilon \, \| \vec{b} - \vec{A}\vec{x}_\star \|.
\end{equation*}
\end{theorem}
In the sections that follow, we will provide an overview of the components that go into \cref{thm:main}.

\subsection{Sparse sign sketch}
\label{sec:sparse_sign}

\begin{definition}\label{def:sparse_sign}
We say $\vec{S}\in\mathbb{R}^{d\times m}$ is a \emph{sparse sign} sketching matrix with sparsity parameter $\zeta$ if
\begin{equation*}
\vec{S} = \sqrt{\frac{m}{\zeta}} \begin{bmatrix}
    \vec{s}_1 & \vec{s}_2 & \cdots & \vec{s}_m
\end{bmatrix},
\end{equation*}
where each column $\vec{s}_i$ is independent and consists of exactly $\zeta$ random signs situated in uniformly random coordinates.
\end{definition}

Sparse sign sketches are efficient to generate and apply
In particular, $\vec{S}$ can be generated in $O(\zeta m)$ time and $\vec{S}\vec{A}$ can be computed in $O(\zeta\nnz(\vec{A}))$ time.
This is substantially faster than $O(dm)$ and $O(d\nnz(\vec{A}))$ times repsectively required to generate and apply a dense sketching matrix.
Moreover, since the columns of $\vec{S}$ are independent, the sketch $\vec{S}$ can easily be generated and applied in parallel.
This is particularly relevant to our setting, since GPUs are particularly well suited for such tasks.
We provide a detailed discussion on implementation in \cref{sec:our_implementation:sparse}.

Sparse sign sketches also produce a good preconditioner within the sketch-and-precondition paradigm. 
In order to quantify the quality of a sketch, it is standard to use the notion of \emph{subspace embedding}.
\begin{definition}\label{def:subspace_embedding}
Let $V$ be a subspace.
We say $\vec{S}$ is an subspace embedding with distortion $\eta$ for $V$ if
\begin{equation*}
    \forall \vec{z}\in V:
\quad
(1-\eta) \|\vec{z}\| \leq \|\vec{S}\vec{z}\| \leq (1+\eta) \|\vec{z}\|.
\end{equation*}
When $V$ is the range of some matrix $\vec{M}$ we say ``subspace embedding for $\vec{M}$''.
\end{definition}

Standard manipulations (see \cref{sec:distortion_and_svals}) reveal that a constant-factor subspace embedding (e.g. $\eta = 1/2$ or $\eta=1/10$) yields an excellent preconditioner and a good initial guess for an iterative method.
\begin{lemma}\label{thm:subspace_embedding_consequences}
If $\vec{S}$ is a subspace embedding for $\vec{A}$ with distortion $\eta$. Then,
\begin{equation*}
    \cond(\vec{A}\vec{M}) \leq \frac{1+\eta}{1-\eta}
\end{equation*}
where $\vec{M}$ is such that $\vec{S}\vec{A}\vec{M}$ has orthonormal columns.
Moreover, if $\vec{S}$ is a subspace embedding for $(\vec{A},\vec{b})$ with distortion $\eta$, then
\begin{equation*}
    \| \vec{A}(\vec{x}_\star - \vec{x}_0) \|
\leq \left( \frac{2\eta}{1-\eta} \right)^{1/2} \| \vec{b} - \vec{A}\vec{x}_\star \|,
\end{equation*}
where $\vec{x}_0$ is the solution to the sketched problem $\min_{\vec{x}}\|\vec{S}\vec{b} - \vec{S}\vec{A}\vec{x}\|$.
\end{lemma}

\begin{remark}
Whether $\vec{S}$  is a subspace embedding for $\vec{A}$ depends only on the range of $\vec{A}$. 
In particular, the conditioning of $\vec{A}$ plays no role!
In the context of iterative methods for \cref{eqn:lstsq} this is remarkable, as it means that the quality of our preconditioner is independent of the conditioning of $\vec{A}$. 
\end{remark}

Sparse sign sketches are known to give a subspace embedding for any fixed problem.
\begin{proposition}[{\protect\cite[Theorem 4.2]{cohen_15}}]\label{thm:cohen}
    For any $(\vec{A},\vec{b})$, a sparse sign sketch is a subspace embedding with distortion $\eta$ with high probability if $d = O(n\log(n)/\eta^2)$ and $\zeta = O(\log(n)/\eta)$.
\end{proposition}

Critically, to ensure \cref{thm:main}, we will only need $\vec{S}$ to be a subspace embedding for $(\vec{A},\vec{b})$ with constant distortion.
Thus, the polynomial dependence on $\eta$ is not of major concern.
Moreover, it is observed numerically that in many cases, $d = O(n)$ and $\zeta = O(1)$ suffice \cite{tropp_yurtsever_udell_cevher_19,martinsson_tropp_20,epperly_23which,dong_martinsson_23}. 
We explore the choices of parameters more in \cref{sec:numerical}.

\subsection{LSQR}
\label{sec:LSQR}

LSQR (\cref{alg:LSQR}) is an iterative method for solving \cref{eqn:lstsq_prec}.
As with most KSMs, the dominant cost of LSQR are matrix-vector products with $\vec{M}$, $\vec{A}$, $\vec{A}^\T$, and $\vec{M}^\T$.
This requires $O(\nnz(\vec{A}) + n^2)\leq O(mn)$ arithmetic operations per iteration.
LSQR is, in a certain sense, optimal among all KSMs.\footnote{LSQR is a particular implementation of the well-known conjugate gradient algorithm applied to the normal equations \cite{paige_saunders_82}.}
In particular the error $\| \vec{A}(\vec{x}_\star - \vec{x}_t)\|$ of the LSQR iterate $\vec{x}_t$ is smaller than for any other KSM using the same initial guess and number of iterations.
This optimality guarantee is sufficient to obtain a convergence guarantee; see \cref{sec:KSMs}.
\begin{proposition}
\label{thm:CG_bound}
The iterate $\vec{x}_t$ produced after $t$ iterations of LSQR with initial guess $\vec{x}_0$ satisfies
\begin{equation*}
\| \vec{A}(\vec{x}_\star - \vec{x}_t) \|
\leq 2 \left( \frac{\cond(\vec{A}\vec{M})-1}{\cond(\vec{A}\vec{M})+1} \right)^t \| \vec{A}(\vec{x}_\star - \vec{x}_0) \|
% \leq 2 \exp\left( -\frac{2t}{\cond(\vec{A}\vec{M})} \right) \| \vec{A}(\vec{x}_\star - \vec{x}_0) \|
.
\end{equation*}
\end{proposition}

Thus, if $\vec{S}$ is a subspace embedding for $(\vec{A},\vec{b})$ with distortion $\eta=1/2$ and we use $\vec{x}_0 = \argmin_{\vec{x}}\|\vec{S}\vec{b} - \vec{S}\vec{A}\vec{x}\|$ as an initial guess and $\vec{M}$ as a preconditioner, 
then \cref{thm:subspace_embedding_consequences} guarantees that
\begin{equation}
t = O(1)\log(1/\varepsilon) 
\quad\Longrightarrow\quad
\| \vec{A}(\vec{x}_\star - \vec{x}_t) \|
\leq \varepsilon  \| \vec{b} - \vec{A}\vec{x}_\star  \|.
\end{equation}
Thus, in combination with \cref{thm:cohen}, which asserts that a sparse sign sketch $d = \tilde{O}(n)$ and $\zeta = \tilde{O}(1)$ gives a subspace embedding with constant distortion, we get \cref{thm:main}.

\begin{algorithm}[ht!]
\caption{Preconditioned LSQR \cite{paige_saunders_82}}
\label{alg:LSQR}
	\begin{algorithmic}[1]
		\Require Matrix $\vec{A}\in\mathbb{R}^{m\times n}$, preconditioner $\vec{M}\in\mathbb{R}^{n\times n}$, vector $\vec{b}\in\mathbb{R}^m$, initial guess $\vec{x}_0\in\mathbb{R}^n$, accuracy target $\varepsilon$
        \State $\beta_1 \vec{u}_1 = \vec{b}-\vec{A}\vec{x}_0$, $\alpha_1 \vec{v}_1 = \vec{M}^\T\vec{A}^\T\vec{u}_1$, $\vec{w}_1 = \vec{M}\vec{v}_1$, $\bar{\phi}_1 = \beta_1$, $\bar{\rho}_1 = \alpha_1$
        \For{$t=1,2,\ldots, $}
        \State $\hat{\vec{u}}_{t+1} = \vec{A}\vec{M}\vec{v}_{t} - \alpha_{t} \vec{u}_{t}$
        \State $\vec{u}_{t+1} = \hat{\vec{u}}_{t+1} / \beta_{t+1}$,~ $\beta_{t+1} = \| \hat{\vec{u}}_{t+1}\|$
        \State $\hat{\vec{v}}_{t+1} = \vec{M}^\T\vec{A}^\T\vec{u}_{t+1} - \beta_{t+1}\vec{v}_{t}$
        \State $\vec{v}_{t+1} = \hat{\vec{v}}_{t+1} / \alpha_{t+1} $,~ $\alpha_{t+1} = \| \hat{\vec{v}}_{t+1}\|$ 
        \State $\rho_{t} = (\bar{\rho}_{t}^2 + \beta_{t+1}^2)^{1/2}$,~  $c_{t} = \bar{\rho}_{t} / \rho_{t}$,~ $s_{t} = \beta_{t+1} / \rho_{t}$,  
        \Statex \hspace{3em}$\theta_{t+1} = s_{t} \alpha_{t+1}$,~ $\bar{\rho}_{t+1} = -c_{t} \alpha_{t+1}$,~ $\phi_{t} = c_{t}\bar{\phi}_{t}$,~ $\bar{\phi}_{t+1} = s_{t} \bar{\phi}_{t}$
        \State $\vec{x}_{t} = \vec{x}_{t-1} + (\phi_{t}/\rho_{t}) \vec{w}_{t}$
        \State $\vec{w}_{t+1} = \vec{M}\vec{v}_{t+1} - (\theta_{t+1}/\rho_t)\vec{w}_{t}$
        \State Test for convergence and break when conditions are met
        \EndFor
	\Ensure Approximate solution $\widehat{\vec{x}} = \vec{x}_{t}$ to \cref{eqn:lstsq}
	\end{algorithmic}
\end{algorithm}

\subsection{Outline and contributions}

In \cref{sec:our_implementation}, we provide a detailed description of our implementation of sketch-and-precondition. 
In particular, we describe several approaches to generating sparse sign sketches.
As there is relatively little documentation on the design choices behind existing implementations of algorithms for generating sparse sign sketches, we believe our discussion will help guide future implementation.

In \cref{sec:numerical} we proceed to benchmark the performance of sketch-and-precondition with sparse sign sketches and LSQR.
These include numerical experiments studying the distortion of sparse sign sketches, which are among the most comprehensive to date (for any computing environment).
We ultimately conclude that sparse sign sketches are efficient to generate and apply, particularly in GPU enabled settings. 
We then study how the choice of the sketch embedding dimension impacts the overall runtime of the sketch-and-precondition and provide some heuristics for setting this parameter in practice. 
Finally, we benchmark the scaling behavior of the algorithm precondition on a (multi)-GPU system.
Our experiments indicate that that our implementation scales reasonably well. 
In particular, our strong-scaling experiments (up to 8 GPUs) never saturate, and our weak scaling experiments indicate that while there is some communication overhead associated with using multiple GPU devices, this cost is mostly overlapped when more than 2 devices are used.

Further numerical experiments, including experiments which justify our choice to focus on sparse sign sketches and LSQR are included in \cref{sec:which_sketch,sec:KSMs} respectively.

% While we focus on GPU-enabled machines, we believe many of our findings naturally extend to other computational environments including standard CPU systems.
% Likewise, our scaling experiments are likely relevant to other distributed memory systems such as as multi-node clusters.

\section{Design and implementation details}
\label{sec:our_implementation}

The aim of this section is to describe, at a high level, some of the design choices we make in our implementation of sketch-and-precondition. 
Our implementation makes use of CuPy \cite{cupy_17}, a  NumPy/SciPy-compatible array library provides a Python interface for key CUDA libraries such as cuBLAS, cuRAND, cuSPARSE, and cuFFT \cite{cuda}.
However, we expect many of the ideas will extend to other software stacks.

\subsection{Generating sparse sign sketches}
\label{sec:our_implementation:sparse}

In this section, we discuss approaches to generating a sparse sign sketch $\vec{S}\in\mathbb{R}^{d\times m}$ (see \cref{sec:sparse_sign}) with sparsity parameter $\zeta$.
As is typical, we store $\vec{S}$ using the standard compressed sparse column (CSC) matrix format, which lets us easily make use of the cuSPARSE sparse matrix multiplication library to compute $\vec{S}\vec{A}$.
Thus, the main implementation task is generating $\vec{S}$.

It is straightforward to sample the values of the $\zeta m$ nonzero entries of $\vec{S}$, which are independent random (scaled) signs.
The more involved task is determining, in each column, which of the $\zeta$ rows will be nonzero.
This task is equivalent to sampling a $\zeta \times m$ matrix 
\begin{equation}
    \vec{C} = 
\begin{bmatrix}
c_{1,1} & c_{1,2} & \cdots & c_{1,m} \\
\vdots & \vdots && \vdots \\
c_{\zeta,1} & c_{\zeta,2} & \cdots & c_{\zeta,m} \\
\end{bmatrix},
\end{equation}
where each column of $\vec{C}$ independently contains exactly $\zeta$ numbers drawn from $[d] = \{1,\ldots, d\}$ \emph{without replacement}.
The CSC representation can then easily be constructed.

\subsubsection{Fisher--Yates type shuffles}

Sampling $\zeta$ numbers uniformly from $[d]$ without replacement is equivalent to taking the first $\zeta$ entries of a uniformly random permutation of $[d]$.\footnote{A number of high-level libraries do sampling without replacement by generating an entire permutation of $[d]$ and then taking the first $\zeta$ entries. This is unnecessarily slow when $\zeta \ll d$, so users must use caution!}
The Fisher--Yates shuffle is an efficient algorithm for this task that avoids generating the entire permutation \cite{knuth_97}.
Given a sample $C$ of $j$ indices drawn from $[d]$ without replacement, the algorithm samples a number from $[d]\setminus C$ uniformly at random as follows:
\begin{enumerate}[noitemsep,topsep=-.5em,label=(\roman*)]
    \item sample an integer $x$ uniformly from $[d-j]$
    \item let $\delta$ be the number of numbers in $C$  less than or equal to $x$
    \item append $x + \delta$ to $C$
\end{enumerate}
The straightforward implementation, where $\delta$ is computed by direct comparison to all numbers already in $C$, requires $O(\zeta^2)$ work per column,\footnote{We assume generating a random integer is unit cost.} and is easily parallelized across columns.
Variants that make use of more complicated data-structures or sorting methods to speed up the comparison cost are also possible. 

The work per column can easily be reduced to $O(\zeta)$ using a standard variant of the shuffle: to avoid comparisons in step (ii), one can maintain a list of $d$ integers, and swap selected with indices from the end of the list.
This is the approach used in RandLAPACK \cite[\S{}A.2.1]{muarray_etal_23}. 
Computing all of $\vec{C}$ with this approach, uses $O(d+m\zeta) = O(m\zeta)$ operations and $O(d + m\zeta) = O(m\zeta)$ storage, which is optimal. 
Parallelization on individual columns is no longer efficient (due to the need to generate a length $d$ list of integers), but parallelization on groups of columns is possible.

\subsubsection{Our approach: rejection sampling}

We make use of an alternate method based on rejection sampling. 
We believe our approach will be preferable to Fisher--Yates type methods in many settings when sparse sign sketches are used, including sketch-and-precondition.
There are a number of ways to implement rejection sampling, the following is a simple approach:
\begin{enumerate}[noitemsep,topsep=-.5em,label=(\roman*)]
    \item Sample a list $C$ of $\zeta$ integers from $[d]$ with replacement
    \item sort $C$ and compare neighbors to identify duplicate entries
    \item replace duplicates with new independent samples drawn uniformly at random from $[d]$
    \item repeat (ii) and (iii) until all entries are unique.
\end{enumerate}

Since the distribution of the number of repeated entries in the first draw has an extremely light tail, the above approach of resampling only offending entries is much more efficient than resampling all entries.
However, even if we needed to resampled the entire column, rejection sampling would still be efficient.
In particular, let $p$ be the probability that a sample of $\zeta$ numbers drawn uniformly at random from $[d]$ with replacement contains no repeats.
We can bound
\begin{equation}
p = \frac{d}{d}\frac{d-1}{d}\cdots\frac{d-(\zeta-1)}{d}
\geq
\bigg(1-\frac{\zeta}{d}\bigg)^\zeta
\geq
1 - \frac{\zeta^2}{d}.
\end{equation}
The theoretical guarantees for sparse sign sketches require $\zeta = O(\log(n))$ and $d = O(n\log(n))$, and $\zeta = O(1)$ and $d = O(n)$ is often used in practice.
In such cases, $p = 1 - \tilde{O}(n^{-1})$, so since $n\gg 1$, rejection sampling is efficient.

A significant practical benefit of our rejection-sampling based approach is that it allows us to leverage high-level primitives such as sorting and array-wise comparison.
As such, this approach is naturally able to take advantage of any system-specific optimizations from existing libraries.
Moreover, these primatives are extremely well-suited for GPUs.
A basic CuPy/SciPy implementation of this approach is given in \cref{alg:sparse_sketch_python}.

\begin{algorithm}[ht]
\caption{CuPy/SciPy code to generate sparse sketching matrix}\label{alg:sparse_sketch_python}
\begin{lstlisting}
def sparse_sketch(d,m,zeta):

    # generate random row-indices with replacement
    C = random.randint(0,d,size=(m,zeta))
    
    # sort indices within columns
    C.sort(axis=1)
    
    # search for duplicates within columns and identify bad columns
    bad_idxs_i,bad_idxs_j = where(C[:,1:] == C[:,:-1])
    n_bad_idxs = len(bad_idxs_j)
    bad_cols = unique(bad_idxs_i)

    # rejection sampling
    while n_bad_idxs > 0:

        # regenerate random row-indices for duplicates
        C[bad_idxs_i,bad_idxs_j] = random.randint(d,size=(n_bad_idxs))
        
        # sort indices (but only in bad columns)
        C[bad_cols] = sort(C[bad_cols],axis=1)
        
        # search for duplicates (but only in bad columns)
        bad_idxs_i,bad_idxs_j = where(C[bad_cols,1:] == C[bad_cols,:-1])
        bad_idxs_i = bad_cols[bad_idxs_i] # convert to global indices
        n_bad_idxs = len(bad_idxs_j)
        bad_cols = unique(bad_idxs_i)

    # random (normalized) signs
    values = ((2/sqrt(zeta))*random.randint(2,size=m*zeta)-1/sqrt(zeta))
    
    # build CSC matrix
    indices = C.flatten()
    indptr = arange(0,m+1)*zeta
    S = sparse.csc_matrix((values,indices,indptr),shape=(d,m))
    
    return S
\end{lstlisting}
\end{algorithm}

\subsection{Preconditioner generate and apply}

The preconditioner $\vec{M}$ is the inverse of a upper triangular matrix $\vec{R}$.
Thus, the preconditioner apply $\vec{x} \mapsto \vec{M}\vec{x}$ can be computed by solving a linear system with $\vec{R}$, an $O(n^2)$ operation.
We instead explicitly form $\vec{M} = \vec{R}^{-1}$ by inverting $\vec{R}$.
This incurs a single $O(n^3)$ cost, and subsequent applies are then $O(n^2)$. 
However, since matrix-vector products are faster than back-substitution on most systems, then we can amortize the $O(n^3)$ of building the inverse over multiple precondition applies.

\subsection{Parallelization on multiple GPUs}
\label{sec:our_implementation:parallel}
In our setting $\vec{A}\in\mathbb{R}^{m\times n}$ is extremely large and tall ($m\gg n \gg 1$).
As such, algorithms are naturally parallelized by distributing any operations involving length-$m$ vectors.
In particular, given $p$ GPU devices, we partition $\{1,2,\ldots, m\}$ into sets of indices $I_1, I_2, \ldots, I_p$ and distribute the corresponding rows of (tall) matrices and vectors to each GPU device. 
We work in a regime where the entire problem can be distributed once and the corresponding parts stored in local GPU memory.
In this setting, many vector operations such as addition or scalar multiplication are efficient and can be done entirely locally.
Operations involving $\vec{A}$ such as matrix-vector products require communication between devices, but the majority of arithmetic work can be done locally.
More details are given in \cref{sec:appendix:our_implementation:parallel}.

\section{Numerical Experiments}
\label{sec:numerical}

We now perform numerical experiments designed to explore the behavior of the implementation described in \cref{sec:our_implementation}.
In \cref{sec:data} we describe our computational environment and the test data we use.
In \cref{sec:sketch_quality} we study how the embedding dimension and sparsity parameter impact the quality of the sketch and the time to build the sketch. 
Then, in \cref{sec:embedding_dim} we observe the impact on the total runtime, and consider a simple heuristic for setting the embedding dimension.
Finally, in \cref{sec:scaling} we study the scaling of our implementation on a multiple GPU system.
Further experiments studying the choice of sketching distribution and iterative method are given in \cref{sec:which_sketch} and \cref{sec:KSMs} respectively.

\subsection{Environment and test problems}
\label{sec:data}

Our experiments are run on a single Intel(R) Xeon(R) Platinum 8275CL compute node equipped with 8 NVIDIA A100-SXM4-40GB GPUs and implemented with CuPy 13.3.0 linked to CUDA 12.8.

% \Tyler{
% Code to reproduce our experiments, along with a more detailed environment description, are available online at \url{url to website ??}.
% }

\begin{table}[]
\centering
\begin{tabular}{ccccc}\toprule
    name & $m$ &$n$ &$\nnz/(mn)$&coherence \\ \midrule
\texttt{5e5\_500\_dense} & $500000$ & $500$ & $1$ & \includegraphics[scale=.5]{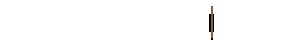}\\
\texttt{5e5\_500\_sparse} & $500000$ & $500$ & $1.0\times 10^{-2}$ & \includegraphics[scale=.5]{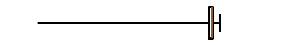}\\
\texttt{5e5\_500\_identity} & $500000$ & $500$ & $2.0\times 10^{-6}$ & \includegraphics[scale=.5]{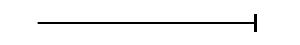}\\
\texttt{MNIST\_standard} & $60000$ & $784$ & $1$ & \includegraphics[scale=.5]{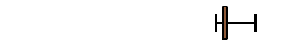}\\
\texttt{bibd\_22\_8} & $319770$ & $231$ & $1.2\times 10^{-1}$ & \includegraphics[scale=.5]{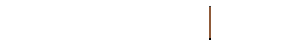}\\
\texttt{nw14} & $123409$ & $73$ & $1.0\times 10^{-1}$ & \includegraphics[scale=.5]{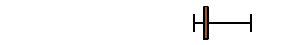}\\
\texttt{connectus} & $394792$ & $512$ & $5.6\times 10^{-3}$ & \includegraphics[scale=.5]{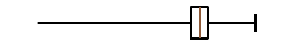}\\
\texttt{rail507} & $63516$ & $507$ & $1.3\times 10^{-2}$ & \includegraphics[scale=.5]{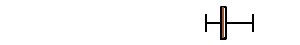}\\
\texttt{pre2\_krylov} & $659033$ & $501$ & $1$ & \includegraphics[scale=.5]{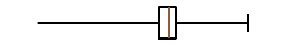}\\
\texttt{t2em\_krylov} & $921632$ & $501$ & $1$ & \includegraphics[scale=.5]{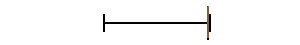}\\
\texttt{parabolic\_fem\_krylov} & $525825$ & $501$ & $1$ & \includegraphics[scale=.5]{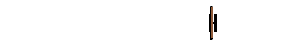}\\
\texttt{kkt\_power\_krylov} & $2063494$ & $501$ & $1$ & \includegraphics[scale=.5]{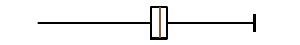}\\
\bottomrule
    \end{tabular}
    \caption{
    Key summary statistics for the test problems used.
    The ``coherence'' column shows a box-and-whiskers plot for the distribution $\{ \| \vec{U}[i] \|^2 : i=1,\ldots,m\}$ on a log-scale ranging from $10^{-15}$ to $1$, where $\vec{U}$ is any orthonormal basis for $\operatorname{range}(\vec{A})$.
    }
    \label{tab:data}
\end{table}

Key summary statistics for the test matrices we used is shown in \cref{tab:data}.
We include several synthetic matrices ($m=500000$ and $n=500$) which are used to probe particular behaviors of sparse sign sketches and sketch-and-precondition.
The first, \texttt{5e5\_500\_identity}, is included specifically as a hard case for sparse sign sketches. 
This matrix consists of the first $n$ columns of the $m\times m$ identity matrix.
The second, \texttt{5e5\_500\_dense}, is a dense matrix with condition number 1000 and left singular vectors roughly drawn from Haar distribution.
The final, \texttt{5e5\_500\_sparse}, is a sparse matrix where each entry is independently nonzero with probability $1/100$, and the value of nonzero entry is a random sign. 
We also include a number of real-world test problems.
The first, is the MNIST training dataset (normalized so that each column has mean zero and variance one) \cite{deng_12}.
We also include several sparse matrices from the SuiteSparse Matrix Collection \cite{suite_sparse_11,suite_sparse_19}.
Finally, we provide several test problem where $\vec{A}$ is the basis for a Krylov subspace generated with a large-sparse square matrix (also from the SparseSuite Matrix Collection).
Such regression problems arise as sub-routine within a modern family of randomized KSMs for large sparse square linear systems \cite[etc.]{balabanov_grigori_22,nakatsukasa_tropp_24,landman_brown_chung_nagy_25}.

Unless indicated otherwise, the curves for each experiment show the median of 100 independent runs of the algorithm, and error bars indicate the 5-95\% range of these trials.
The test problems remain fixed across these trials.

\subsection{Sketch parameters}
\label{sec:sketch_quality}

Sparse sign sketches are parameterized by the sparsity parameter $\zeta$ and embedding dimension $d$.
These parameters impact how fast the sketching matrix can be generated and applied, as well as the quality of the sketch produced.

\subsubsection{Sketch generate and apply time}

\begin{figure}[ht]
    \centering
    \includegraphics[scale=0.7]{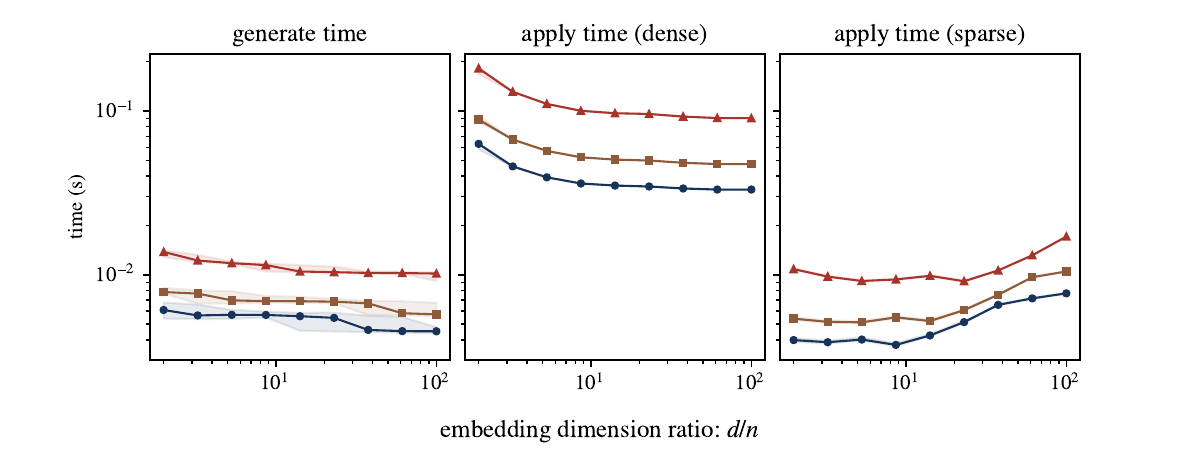}
    \caption{Time to generate and apply the sparse sign sketching matrix as a function of embedding dimension, for several choices of sparsity: $\zeta=8$  ({\protect\raisebox{.25pt}{\protect\includegraphics[scale=0.85]{imgs/legend/l0.pdf}}}), $\zeta=12$ ({\protect\raisebox{.25pt}{\protect\includegraphics[scale=0.85]{imgs/legend/l1.pdf}}}), $\zeta=24$ ({\protect\raisebox{.25pt}{\protect\includegraphics[scale=0.85]{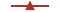}}}).
    }
    \label{fig:ed_sparsity_time}
\end{figure}

In \cref{fig:ed_sparsity_time} we compare the time required to generate and apply sparse sign sketches with various choices of $\zeta$ and $d$.
In all cases, we observe the cost is essentially proportional to the sparsity parameter $\zeta$; i.e., that doubling $\zeta$ doubles the cost.
We observe that the time to generate the sketch decreases slightly as the embedding dimension increases. 
This is the expected behavior, as rejection sampling succeeds more frequently; see \cref{sec:our_implementation:sparse}).
When applying $\vec{S}$ to a \texttt{5e5\_500\_dense}, which is a dense matrix, we observe that the cost also decreases some as the embedding dimension increases, a result of reduced overhead from the sparse matrix library.
On the other hand, when applying $\vec{S}$ to \texttt{5e5\_500\_sparse}, where $\nnz(\vec{A})/(mn) \approx 0.01$, the apply time is much lower.
However, the apply time increases some as the embedding dimension increases, presumably due to overhead in the sparse matrix multiplication algorithm.

\subsubsection{Distortion}
\label{sec:ed_sparsity_distortion}

\begin{figure}[ht]
    \centering
    \includegraphics[scale=0.7]{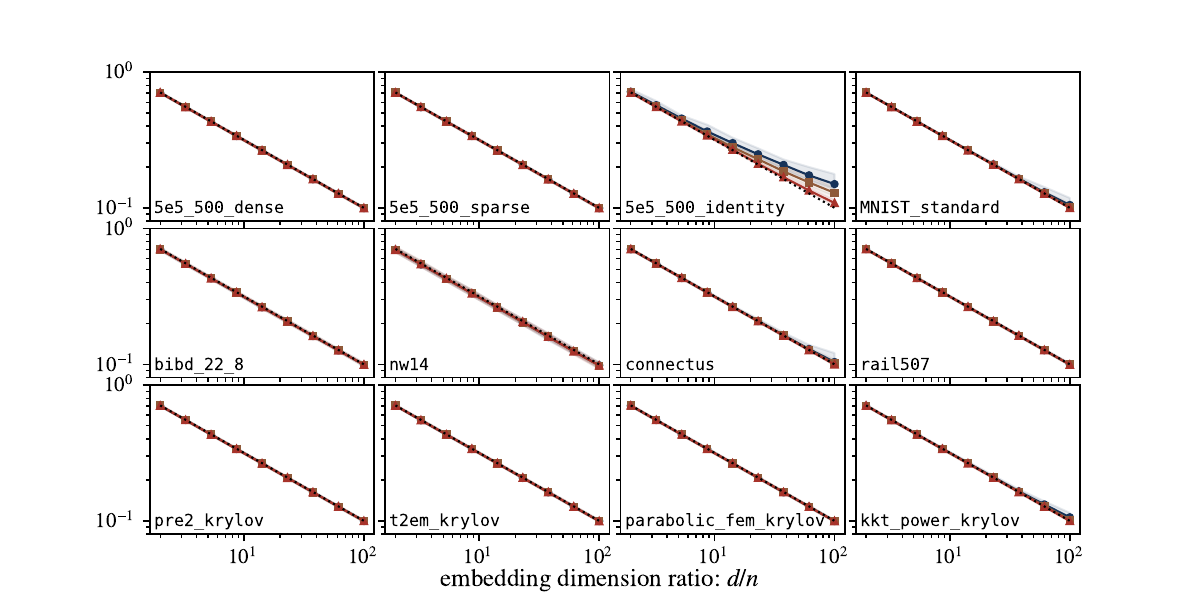}
    \caption{
    Distortion as a function of relative embedding dimension $d/n$ for sparse sign sketches with several choices of sparsity: $\zeta=8$  ({\protect\raisebox{.25pt}{\protect\includegraphics[scale=0.85]{imgs/legend/l0.pdf}}}), $\zeta=12$ ({\protect\raisebox{.25pt}{\protect\includegraphics[scale=0.85]{imgs/legend/l1.pdf}}}), $\zeta=24$ ({\protect\raisebox{.25pt}{\protect\includegraphics[scale=0.85]{imgs/legend/l2.pdf}}}).
    Note that sparse sketches behave remarkably similarly to the asymptotic theory for Gaussian sketches which have distortion $\sqrt{n/d}$ ({\protect\raisebox{.25pt}{\protect\includegraphics[scale=0.85]{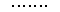}}}).
    }
    \label{fig:ed_sparsity_distortion}
\end{figure}

Next, we study how the distortion of the sketch (see \cref{def:subspace_embedding}) changes with the parameters $\zeta$ and $d$.
In \cref{fig:ed_sparsity_distortion} we plot the distortion as a function of sparsity and embedding dimension for several test problems. 
For reference, we also plot the reference line $\sqrt{n/d}$, which is the asymptotic distortion $\sqrt{n/d}$ of a Gaussian sketch when $n,d\to\infty$ with $d/n$ held constant \cite{martinsson_tropp_20}.

In large, the performance of the sparse sign sketch seems to  follow the Gaussian theory, even with $\zeta=8$.
In fact, the only example with a notable deviation from the Gaussian theory is the synthetic example \texttt{5e5\_500\_identity}, which was included specifically as a hard case for sparse sign sketches.\footnote{This example is used in \cite{nelson_nguyen_14} to prove lower bounds for sparse sketching distributions.}
On this example, we see that using small values of $\zeta$ causes some problems when the embedding dimension is large.
This can be remedied by increasing the sparsity parameter $\zeta$.

\subsection{Embedding dimension}
\label{sec:embedding_dim}

As observed in \cref{fig:ed_sparsity_time} and reflected by the theory in \cref{sec:sparse_sign}, the time needed to generate and apply sparse sketches is roughly independent of the embedding dimension $d$.
Thus, we might be tempted to use a large embedding dimension in order to improve the convergence of LQSR.
However, the larger the embedding dimension, the longer the QR factorization in \cref{alg:main:qr} of \cref{alg:main} will take; in particular, the runtime of this step is $O(dn^2)$.
Thus, the increased time of this step must be offset by any improvement to the convergence of LSQR resulting from a smaller distortion.\footnote{A smaller distortion also results in a better initial guess, but this impact is relatively small compared to the improvements due to the faster rate of convergence.}

Remarkably (although unsurprisingly in light of the experiments in \cref{sec:sketch_quality}), we can often estimate the rate of convergence of preconditioned LSQR quite accurately.
In particular, the rate 
\begin{equation}\label{eqn:estimate_rate}
(n/d)^{t/2} = \left( \frac{\kappa-1}{\kappa+1} \right)^t
,\qquad
\kappa = \frac{1+\sqrt{n/d}}{1-\sqrt{n/d}}
\end{equation}
typically gives a good estimate of the true convergence.
This estimate is obtained by using the convergence bound \cref{thm:CG_bound} for LSQR, and the condition number bound $\cond(\vec{A}\vec{M}) \leq (1+\eta)/(1-\eta)$, and then estimating $\eta \approx \sqrt{n/d}$.

\subsubsection{Impact on runtime}

\begin{figure}[ht]
    \centering
    \includegraphics[scale=.7]{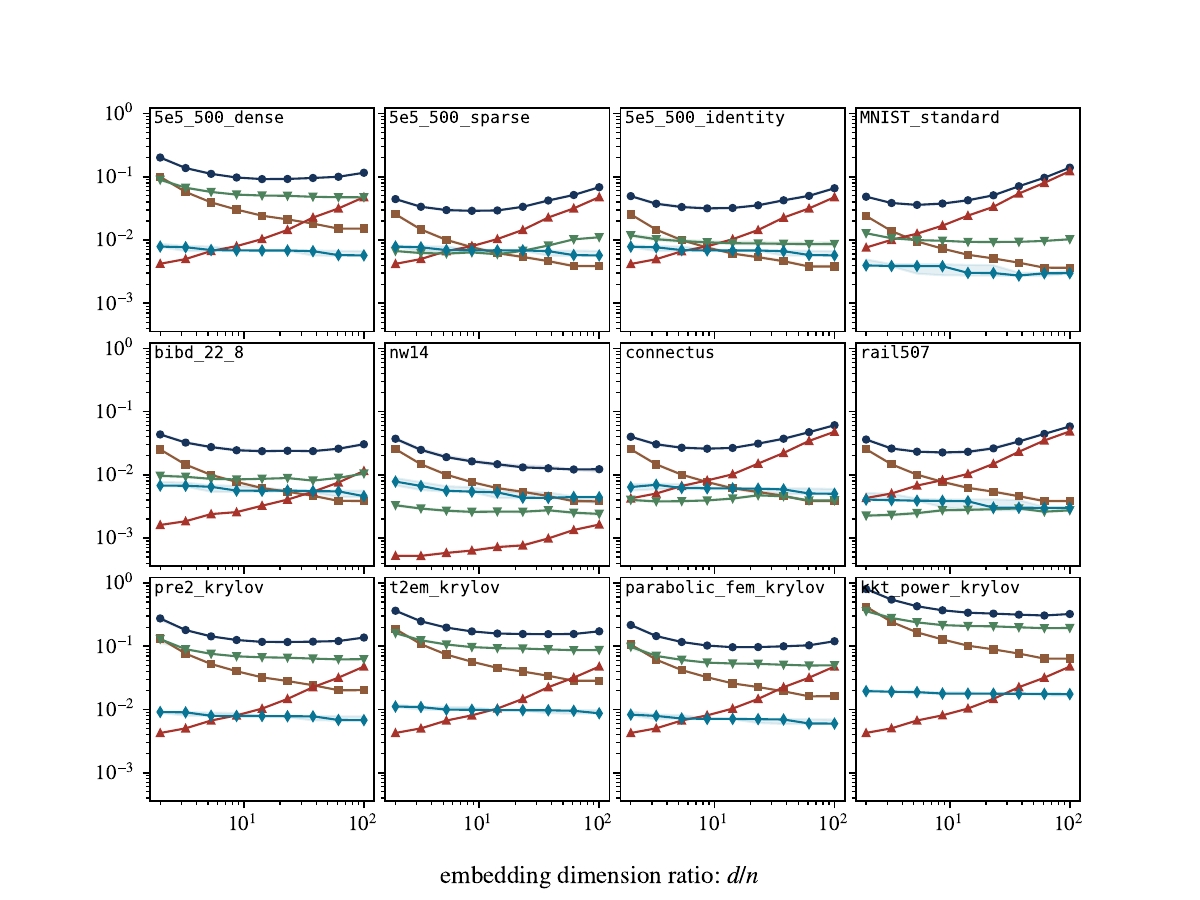}
    \caption{
    Total runtime of sketch-and-precondition ($\zeta=12$) as a function of embedding dimension ({\protect\raisebox{.25pt}{\protect\includegraphics[scale=0.85]{imgs/legend/l0.pdf}}}) and individual components: 
    iteration time ({\protect\raisebox{.25pt}{\protect\includegraphics[scale=0.85]{imgs/legend/l1.pdf}}}),
    preconditioner build time ({\protect\raisebox{.25pt}{\protect\includegraphics[scale=0.85]{imgs/legend/l2.pdf}}}),
    sketch apply time
    ({\protect\raisebox{.25pt}{\protect\includegraphics[scale=0.85]{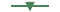}}}), and
    sketch generate time
    ({\protect\raisebox{.25pt}{\protect\includegraphics[scale=0.85]{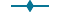}}}).
    }
    \label{fig:embedding_dim_runtime}
\end{figure}

In \cref{fig:embedding_dim_runtime} we compare the runtime of \cref{alg:main} for various embedding dimensions.
The number of iterations is set to $t = \log(10^{-10})/\log(n/d)$ so that LSQR decreases the error by roughly 10 orders of magnitude.
As expected, as the embedding dimension increases, the time spent on iteration decreases, but the time required to build the preconditioner increases. 
Using too small or too large of an embedding dimension  results in a longer overall runtime.

\subsubsection{Selecting the embedding dimension}

The upshot is that \cref{eqn:estimate_rate} allows us to estimate the total number of iterations required to get convergence, using only $n$ and the proposed embedding dimension $d$.
As such, if the preconditioner build time and time per iteration can be estimated, then the embedding dimension can easily be chosen to (approximately) minimize the total runtime.

As a proof of concept, consider the case of a dense $\vec{A}$ where each iteration of LSQR requires  $O(mn)$ arithmetic operations.
We require $t= O(\log(\varepsilon)/\log(n/d))$ to ensure \cref{eqn:estimate_rate} is of size $\varepsilon$.
Building the preconditioner via a QR factorization requires $O(dn^2)$ arithmetic operations.
A direct computation reveals that
\begin{equation}
\label{eqn:ed_balance}
d = n \exp\left(W\left(-\frac{m\log(\varepsilon)}{n^2}\right)\right)
\quad\Longrightarrow\quad
\frac{\log(\varepsilon)}{\log({n/d})}\cdot mn =  dn^2,
\end{equation}
where $W(\,\cdot\,)$ is the Lambert-W function.

\begin{figure}
\centering
\includegraphics[scale=.7]{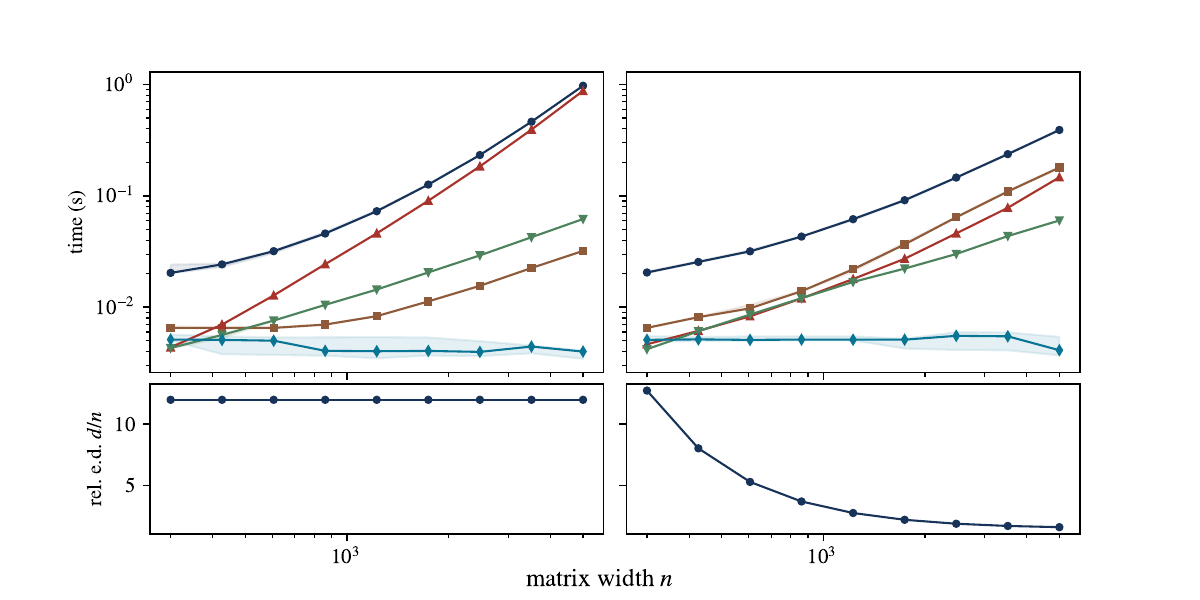}
\caption{
Comparison of procedures for selecting the sketch-and-precondition ($\zeta=12$) embedding dimension for $m=6\times 10^5$ and $n$ ranging from $3\times 10^2$ to $5\times 10^3$: fixed embedding dimension (left) and heuristic optimization \cref{eqn:ed_balance} (right).
The top panels show the total runtime ({\protect\raisebox{.25pt}{\protect\includegraphics[scale=0.85]{imgs/legend/l0.pdf}}}) as a function of the matrix width, as well as individual components: iteration time ({\protect\raisebox{.25pt}{\protect\includegraphics[scale=0.85]{imgs/legend/l1.pdf}}}),
preconditioner build time ({\protect\raisebox{.25pt}{\protect\includegraphics[scale=0.85]{imgs/legend/l2.pdf}}}),
sketch apply time
({\protect\raisebox{.25pt}{\protect\includegraphics[scale=0.85]{imgs/legend/l3.pdf}}}), and
sketch generate time
({\protect\raisebox{.25pt}{\protect\includegraphics[scale=0.85]{imgs/legend/l4.pdf}}}).
The bottom panels show the choice of embedding dimension as a function of the matrix width.
}
\label{fig:embedding_dim_est}
\end{figure}

In \cref{fig:embedding_dim_est} we illustrate the runtime on a sequence of test matrices with height $m=6\times 10^5$ and increasing with $n$ ranging from $3\times 10^2$ to $5\times 10^3$.
For each matrix, a right hand side is generated so that $\|\vec{b}\| = 1$ and $\|\vec{b} - \vec{A}\vec{x}^*\|=1/2$. 
The components of $\vec{b}$ in the range of $\vec{A}$ and orthogonal compliment of the range of $\vec{A}$ are drawn from the uniform distribution (and scaled appropriately).

The left panel shows the costs associated with a constant embedding dimension $d=12n$ (which is suggested in \cite{epperly_meier_nakatsukasa_24}) and the right panel shows the costs associated with an adaptively chosen embedding dimension based on the approach described in \cref{eqn:ed_balance}. 
Here LSQR is run for a number of iterations so that the final error is less than $\|\vec{A}(\vec{x}_\star - \widehat{\vec{x}})\| \leq \varepsilon \| \vec{b} \|$, where $\varepsilon = 10^{-5}$.

We observe the estimate \cref{eqn:ed_balance} results in substantially smaller running times when $n$ is sufficiently large. 

This experiment suggests that a practical implementation of sketch-and-precondition should attempt to balance the costs of the iterative method with the preconditioner build. 
The choice we made in \cref{eqn:ed_balance} is simple, but one could imagine a better balancing optimization based on environment-specific benchmarking.
We leave this to future work.

\subsection{Multi-GPU scaling}
\label{sec:scaling}

We now study the scalability of our implementation of \cref{alg:main} on multiple GPUs. 
We focus on the case that $\vec{A}$ is dense.

\subsubsection{Strong scaling}

We begin with a strong-scaling experiment. 
Here, we fix a problem, and increase the number GPUs available.
In the best case, the runtime will be inversely proportional to the number of GPUs available. 
However, due to non-parallelized parts of the algorithm and overhead associated with communication between GPUs, this ideal scaling rate is not necessarily achieved.

\begin{figure}[htb]
    \centering
    \includegraphics[scale=.7]{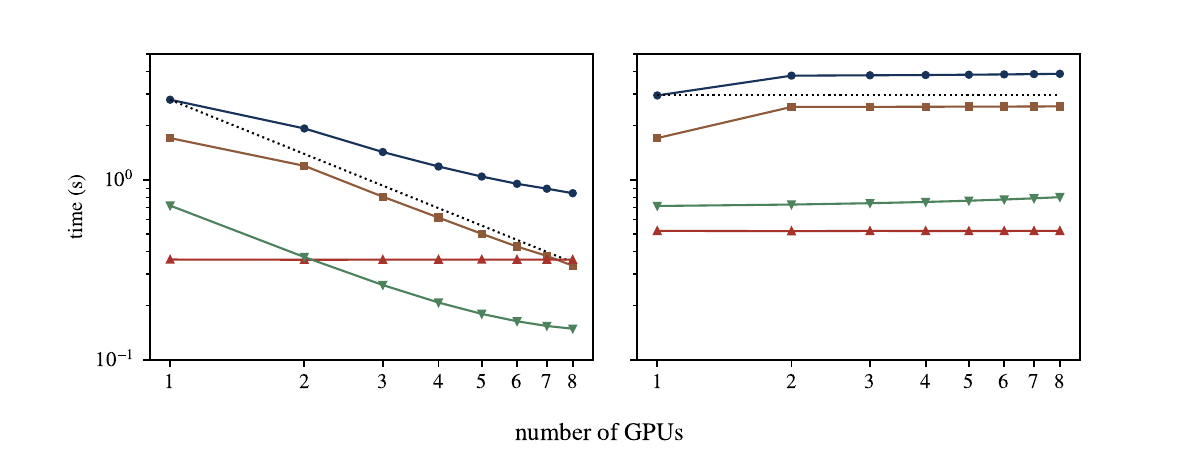}
    \caption{Strong and weak scaling experiment indicating
    total runtime ({\protect\raisebox{.25pt}{\protect\includegraphics[scale=0.85]{imgs/legend/l0.pdf}}}),
    total iteration time ({\protect\raisebox{.25pt}{\protect\includegraphics[scale=0.85]{imgs/legend/l1.pdf}}}), 
    sketch generate/apply time  ({\protect\raisebox{.25pt}{\protect\includegraphics[scale=0.85]{imgs/legend/l3.pdf}}}), 
    and preconditioner build time 
    ({\protect\raisebox{.25pt}{\protect\includegraphics[scale=0.85]{imgs/legend/l2.pdf}}}).
    Ideal scaling rate $\sim\!1/\text{(\# GPUs)}$ for weak scaling and ($\sim\!1$) for strong scaling ({\protect\raisebox{.25pt}{\protect\includegraphics[scale=0.85]{imgs/legend/dotted.pdf}}}).
    }
    \label{fig:scaling}
\end{figure}

In the left panel of \cref{fig:scaling} we show the results of our experiment. 
We use a dense matrix with $m = 10^6$ and $n=4\times 10^3$. 
The embedding dimension is $d = 8n$ and sparsity parameter $\zeta = 12$.
We observe that when moving from 1-2 GPUs, there is an improvement in the runtime, but the improvement is substantially less than the ideal $2\times$ speedup.
However, with 2-8 GPUs, the scaling is much better. 
This indicates that the communication costs are roughly constant regardless of the number of GPU.
However, since the preconditioner build is not parallelized, the total runtime begins to flatten out as the cost of the iterative method, which is parallelized, becomes small relative to the cost ofwow,  the of the preconditioner build.

\subsubsection{Weak scaling}

Next, we perform a weak-scaling experiment, where the problem size is increased proportional to the number of GPUs available.
This more directly allows us to observe the impact of communication.

In the right panel of \cref{fig:scaling} we show the results of our experiment. 
Here we increase the problem size as we increase the number of GPUs available. 
In particular, we form $\vec{A}$ and $\vec{b}$ by stacking copies of $\vec{A}_{\textup{small}}$ and $\vec{b}_{\textup{small}}$, where the number of copies is equal to the number of GPUs available.
Here $\vec{A}_{\textup{small}}$ is the same dense matrix as in the previous section.
Hence, on 8 GPUs, we are solving a problem with $m = 8\times 10^6$ and $n=4\times 10^3$, requiring roughly 256GB of memory to store.
We now use an embedding dimension $d=12n$.
As in the strong scaling experiment, we observe an overhead when moving from 1-2 GPUs, but cost for running with 2-8GPUs remains nearly constant.

\section{Conclusions and future work}

We have described an implementation of sketch-and-precondition with sparse sign sketches and LSQR. 
To the best of our knowledge, our implementation of sparse sign sketches is one of the first efficient high-level implementations.
We hope that our code and the discussion around particular implementation choices will help guide future development of software for sparse sign sketches.
Extensive benchmarking of sparse sign sketches and a sparse sign sketch and LSQR based implementation of sketch-and-precondition confirm that, as predicted by theory, the algorithm works well in practice.

Our numerical experiments also suggest that a practical implementation of sketch-and-precondition should attempt to balance the cost of building the preconditioner and the cost of running a preconditioned iterative method.
While we have demonstrated preliminary results in this direction, it would be interesting to perform system-specific benchmarking in order to optimize this tradeoff.
We are currently unaware of any existing efforts in this directions.

Finally, while we observed numerically that sparse sign sketches behave well in most cases, even if $\zeta = O(1)$, a theoretical understanding for this is missing. 
A theoretical explanation for this phenomenon may provide insight into how the sparsity parameter should be set in practice.

\section{Acknowledgments}

We thank our colleagues at the Global Technology Applied Research center of JPMorganChase for support and helpful feedback. Special thanks to Shouvanik Chakrabarti and Dylan Herman for their valuable discussions regarding the manuscript.

\printbibliography

\vfill
\section*{Disclaimer}

This paper was prepared for informational purposes by the Global Technology Applied Research center of JPMorgan Chase \& Co. This paper is not a merchandisable/sellable product of the Research Department of JPMorgan Chase \& Co. or its affiliates. Neither JPMorgan Chase \& Co. nor any of its affiliates makes any explicit or implied representation or warranty and none of them accept any liability in connection with this paper, including, without limitation, with respect to the completeness, accuracy, or reliability of the information contained herein and the potential legal, compliance, tax, or accounting effects thereof. This document is not intended as investment research or investment advice, or as a recommendation, offer, or solicitation for the purchase or sale of any security, financial instrument, financial product or service, or to be used in any way for evaluating the merits of participating in any transaction.
\clearpage
\appendix

\section{Theoretical properties implied by a subspace embedding}

In this section we provide the manipulations which lead to \cref{thm:subspace_embedding_consequences}.

\subsection{Singular values of the preconditioned problem}
\label{sec:distortion_and_svals}

Suppose $\vec{S}$ is a subspace embedding for $\vec{A}$ with distortion $\eta$; i.e.,
\begin{equation}
    \label{def:subspace_embedding_appendix}
    \forall \vec{z}\in \operatorname{range}(\vec{A}):
\quad
(1-\eta) \|\vec{z}\| \leq \|\vec{S}\vec{z}\| \leq (1+\eta) \|\vec{z}\|.
\end{equation}
For any invertible $\vec{M}$, we can reparameterize $\vec{z} = \vec{A}\vec{M}\vec{c}$ for $\vec{c}\in\mathbb{R}^n$.
Then \cref{def:subspace_embedding_appendix} is equivalent to
\begin{equation}
\label{def:subspace_embedding_appendix_c}
\forall \vec{c}\in\mathbb{R}^n:
\quad
(1-\eta) \|\vec{A}\vec{M}\vec{c}\| \leq \|\vec{S}\vec{A}\vec{M}\vec{c}\| \leq (1+\eta) \|\vec{A}\vec{M}\vec{c}\|.
\end{equation}
When $\vec{S}\vec{A}\vec{M}$ has orthonormal columns,  $\|\vec{S}\vec{A}\vec{M}\vec{c}\| = \|\vec{c}\|$.
Therefore, \cref{def:subspace_embedding_appendix_c} becomes
\begin{equation}\label{def:subspace_embedding_appendix_AMc}
\forall \vec{c}\in\mathbb{R}^n:
\quad
(1-\eta) \frac{\|\vec{A}\vec{M}\vec{c}\|}{\|\vec{c}\|} \leq 1 \leq (1+\eta) \frac{\|\vec{A}\vec{M}\vec{c}\|}{\|\vec{c}\|}.
\end{equation}
Recall 
\begin{equation}
    \smin(\vec{AM}) = \min_{\vec{c}\in\mathbb{R}^n} \frac{\|\vec{A}\vec{M}\vec{c}\|}{\|\vec{c}\|}
    ,\qquad
    \smax(\vec{AM}) = \max_{\vec{c}\in\mathbb{R}^n} \frac{\|\vec{A}\vec{M}\vec{c}\|}{\|\vec{c}\|}.
\end{equation}
Then \cref{def:subspace_embedding_appendix_AMc} becomes
\begin{equation}
    (1-\eta) \smax(\vec{A}\vec{M}) \leq 1 \leq (1+\eta) \smin(\vec{A}\vec{M}),
\end{equation}
and hence 
\begin{equation}
    \cond(\vec{A}\vec{M}) = \frac{\smax(\vec{A}\vec{M})}{\smin(\vec{A}\vec{M})}
    \leq \frac{1+\eta}{1-\eta}.
\end{equation}

\begin{remark}
Note that this entire argument depends only on $\operatorname{range}(\vec{A})$ and not on the singular values or right singular vectors!
\end{remark}

\subsection{Sketch-and-solve}

Suppose $\vec{S}$ is a subspace embedding for $(\vec{A},\vec{b})$ with distortion $\eta$; i.e.,
\begin{equation}
    \forall \vec{z}\in \operatorname{range}(\vec{A})+\operatorname{span}(\vec{b}):
\quad
(1-\eta) \|\vec{z}\| \leq \|\vec{S}\vec{z}\| \leq (1+\eta) \|\vec{z}\|.
\end{equation}
Let $\vec{x}_0 \in \mathbb{R}^n$ be the solution to the sketched problem, i.e.,
\begin{equation}
    \| \vec{S}\vec{b} - \vec{S}\vec{A} \vec{x}_0  \| = \min_{\vec{x}\in\mathbb{R}^n} \|\vec{S}\vec{b} - \vec{S}\vec{A}\vec{x}\|.
\end{equation}
Then, since $\vec{b} - \vec{A}\vec{x} \in \operatorname{range}(\vec{A})+\operatorname{span}(\vec{b})$, using both sides of \cref{def:subspace_embedding},
\begin{equation}
    \| \vec{b} - \vec{A} \vec{x}_0  \| 
    \leq \frac{1}{1-\eta}\|\vec{S}(\vec{b} - \vec{A}\vec{x}_0) \|
    = \frac{1}{1-\eta} \min_{\vec{x}\in\mathbb{R}^n} \|\vec{S}(\vec{b} - \vec{A}\vec{x})\|
    \leq \frac{1+\eta}{1-\eta} \min_{\vec{x}\in\mathbb{R}^n} \|\vec{b} - \vec{A}\vec{x}\|
    =  \frac{1+\eta}{1-\eta} \| \vec{b} - \vec{A} \vec{x}_\star \|.
\end{equation}
So, a subspace embedding with distortion $\eta$ guarantees an approximation with residual error at most $(1+O(\eta))$ times best possible.

\begin{remark}
In fact, a more careful analysis \cite{epperly_25accurate} reveals that in fact a $(1+O(\eta^2))$ relative residual norm is guaranteed.
In particular, 
\begin{equation}
    \| \vec{b} - \vec{A} \vec{x}_0  \| 
    \leq \left( 1 + \frac{(2\eta+\eta^2)^2}{(1-\eta)^4} \right)^{1/2} \| \vec{b} - \vec{A} \vec{x}_\star \|.
\end{equation}
\end{remark}

\subsection{Residual error to forward error}

In \cref{thm:subspace_embedding_consequences} we state guarantees in terms of the forward error. 
As we now show, this is, in a certain sense, equivalent to bound in terms of the residual norm.

The true residual $\vec{b} - \vec{A}\vec{x}_\star$ is orthogonal to the range of $\vec{A}$; i.e., $\vec{A}^\T(\vec{b} - \vec{A}\vec{x}_\star) = \vec{0}$.
Thus, for any $\widehat{\vec{x}} \in \mathbb{R}^n$, by the Pythagorean theorem,
\begin{align}
    \|\vec{b} - \vec{A}\widehat{\vec{x}}\|^2
    &= \| \vec{b} - \vec{A}(\vec{x}_\star + \widehat{\vec{x}}- \vec{x}_\star) \|^2
    \\&= \| \underbrace{\vec{b} - \vec{A}\vec{x}_\star}_{\in \operatorname{span}(\vec{A})^\perp} - \underbrace{\vec{A} (\vec{x}_\star - \widehat{\vec{x}})}_{\in \operatorname{span}(\vec{A})} \|^2
    \\&= \| \vec{b} - \vec{A}\vec{x}_\star \|^2 + \| \vec{A} (\vec{x}_\star - \widehat{\vec{x}})\|^2.
\end{align}
Rearranging, we find that
\begin{equation}
    \| \vec{A}(\vec{x}_\star - \widehat{\vec{x}})\|
    = \big( \|\vec{b} - \vec{A}\widehat{\vec{x}}\|^2 - \| \vec{b} - \vec{A}\vec{x}_\star \|^2 \big)^{1/2}.
\end{equation}

\section{Parallelization on multiple GPUs}
\label{sec:appendix:our_implementation:parallel}

As mentioned in \cref{sec:our_implementation:parallel}, we parallelize by distributing the rows across GPU devices. 
A simplified version of the Python class we use to manage this is described in \cref{alg:dist_array}.
We implement class methods for basic linear algebra tasks such as  scalar multiplication, vector addition, and products with $\vec{A}$ or $\vec{A}^\T$.

Given a index sets $I$ and $J$, we write $\vec{X}[I,J]$ to denote the submatrix of $\vec{X}$ corresponding to the rows and columns in $I$ and $J$ respectively.
If $J$ contains all possible entries, we will write $X[I,\,\cdot\,]$ and similarly $\vec{X}[\,\cdot\,,J]$ if $I$ contains all possible entries.

\begin{algorithm}
\caption{Distributed Array Class}\label{alg:dist_array}
\begin{lstlisting}
class DistributedRowArray():
    
  def __init__(self,data,n_devices):

    # set basic attributes
    self.n_devices = n_devices
    self.dtype = data.dtype
    self.shape = data.shape
    
    # define how matrix will be split
    self.row_split = []
    stride = self.shape[0] // n_devices
    for rank in range(self.n_devices):
      start = rank*stride
      stop = (rank+1)*stride if rank < n_devices-1 else None
      self.row_split.append(slice(start,stop))

    # distribute chunks of data to GPU devices
    self.chunks = [None for rank in range(self.n_devices)]
    for rank in range(self.n_devices):
      with cp.cuda.Device(rank):
        self.chunks[rank] = cp.asarray(data[self.row_split[rank]])  
\end{lstlisting}    
\end{algorithm}

\subsection{Basic linear algebra}

Operations such as vector addition, scalar multiplication, and taking norms are also communication efficient.
Indeed, for vectors $\vec{y},\vec{z}\in\mathbb{R}^m$,
\begin{equation}
\vec{y} + \alpha \vec{z}
= 
\begin{bmatrix}
    \vec{y}[I_0] \\ \vdots \\ \vec{y}[I_p] 
\end{bmatrix}
+ \alpha
\begin{bmatrix}
    \vec{z}[I_0] \\ \vdots \\ \vec{z}[I_p] 
\end{bmatrix}
= 
\begin{bmatrix}
    \vec{y}[I_0] + \alpha \vec{z}[I_0]  \\ \vdots \\ \vec{y}[I_p] + \alpha \vec{z}[I_p] 
\end{bmatrix}
,\qquad
\|\vec{y}\|^2
= \|\vec{y}[I_1]\|^2 + \cdots + \|\vec{y}[I_p]\|^2.
\end{equation}
In particular, vector addition only requires adding the local rows on each GPU device, and scalar multiplication require copying a single scalar to all GPU devices and then performing a local scalar multiplication on the corresponding locally held rows.
Computing the norm requires accumulating all of the row-norms held on each device. 
When appropriate, operations are done in place.

\Cref{alg:inplace_add} illustrates how an in place addition operation for two \texttt{DistributedRowArray} objects can be implemented.
Other operations are implemented analogously.

\begin{algorithm}
\caption{In place addition}\label{alg:inplace_add}
\begin{lstlisting}
def __iadd__(self,y):
    for rank in range(self.n_devices):
        with cp.cuda.Device(rank):
            self.chunks[rank] += y.chunks[rank]
            
    return self
\end{lstlisting}    
\end{algorithm}

Operations with (short) length-$n$ vectors and the products with $\vec{M}$ and $\vec{M}^\T$ are performed on one GPU device, and then communicated to other devices as necessary.

\subsection{Products with the data matrix}

Products with $\vec{A}$ and $\vec{A}^\T$ are also easily parallelized.
Indeed, given $\vec{x}\in\mathbb{R}^n$ ,
\begin{equation}
\vec{A}\vec{x} = 
\begin{bmatrix}
    \vec{A}[I_1,\,\cdot\,] \\
    \vdots \\
    \vec{A}[I_p,\,\cdot\,]
\end{bmatrix}
\vec{x}
=
\begin{bmatrix}
    \vec{A}[I_1,\,\cdot\,]\vec{x} \\
    \vdots \\
    \vec{A}[I_p,\,\cdot\,]\vec{x}
\end{bmatrix}
\end{equation}
and given $\vec{y}\in\mathbb{R}^m$
\begin{equation}
\vec{A}^\T\vec{y} = 
\begin{bmatrix}
    \vec{A}[I_1,\,\cdot\,] \\
    \vdots \\
    \vec{A}[I_p,\,\cdot\,]
\end{bmatrix}^\T
\begin{bmatrix}
    \vec{y}[I_0] \\ \vdots \\ \vec{y}[I_p] 
\end{bmatrix}
= \vec{A}[I_1,\,\cdot\,]^\T\vec{y}[I_1,\,\cdot\,] + \cdots + \vec{A}[I_p,\,\cdot\,]^\T\vec{y}[I_p].
\end{equation}
In particular, since $\vec{A}$ remains fixed throughout, we only need to distribute the relevant rows of $\vec{A}$ to the given GPU device a single time at the outset.

\subsection{Sketching}

For sketching distributions where the columns of $\vec{S}$ are independent, generating $\vec{S}$ and computing $\vec{S}\vec{A}$ is also extremely easy to parallelize since
\begin{equation}
\vec{S} \vec{A} = 
\begin{bmatrix}
    \vec{S}[\,\cdot\,,I_1] & \cdots & \vec{S}[\,\cdot\,,I_p]
\end{bmatrix}
\begin{bmatrix}
    \vec{A}[I_1,\,\cdot\,] \\
    \vdots \\
    \vec{A}[I_p,\,\cdot\,]
\end{bmatrix}
= \vec{S}[\,\cdot\,,I_1]\vec{A}[I_1,\,\cdot\,] + \cdots + \vec{S}[\,\cdot\,,I_p]\vec{A}[I_p,\,\cdot\,].
\end{equation}

\section{Which sketch should I use?}
\label{sec:which_sketch}

While we have focused primarily on sparse sign sketches, as mentioned in \cref{sec:sparse_sign}, there are many other sketching distributions.
Two of the most commonly used are Gaussian (or other dense) sketching matrices and subsampled fast trigonometric transform sketches.
In this section we provide a further comparison with sparse sign sketches.\footnote{This section shares a title with Ethan Epperly's blog \cite{epperly_23which}, which provides many related tests and influenced the design of our experiments.}
The experiments in this section are are intended primarily to illustrate some key tradeoffs between the sketch methods and should not be viewed as a comprehensive or definitive comparison of the methods. 

In our experiments the synthetic matrices are replaced with smaller ($m=200000$ and $n=200$) analogs, \texttt{2e5\_200\_identity}, \texttt{2e5\_200\_dense}, \texttt{2e5\_200\_sparse}, and the Krylov matrices are restricted to the first $n=101$ columns.

\subsection{Other sketches}
\label{sec:other_sketches}

\paragraph{Gaussian sketch}

Perhaps the simplest sketching distribution is if $\vec{S}$ has independent Gaussian entries.
Owing to the Gaussian distribution's integral role in probability and random matrix theory, the theoretical guarantees for Gaussian sketches are simple and strong.
In particular, it suffices to set $d = O(n)$ to ensure a version of \cref{thm:main}; see e.g. \cite[\S8]{martinsson_tropp_20}.
This is the optimal dependence on $n$, as we certainly require $d\geq n$ to ensure $\eta < 1$.

A downside to Gaussian sketches is that their costs depend strongly on the embedding dimension.
In particular, $O(md)$ Gaussian random variables must be generated and applying $\vec{S}$ to $\vec{A}$ requires $O(\nnz(\vec{A})d)$ operations.
If $\vec{A}$ is dense, this is the same arithmetic as the cost of using a direct method to solve \cref{eqn:lstsq} and is often intractable.\footnote{Even in the dense case, sketch-and-precondition with Gaussian sketches still offers potential practical speedups over direct methods.
Indeed, the main cost becomes a large matrix-matrix multiplication, which is well suitable to take advantage of of any parallelism that may be present in a computing environment \cite{meng_saunders_mahoney_14}.}

\paragraph{Subsampled fast trigonometric transform}

To be able to apply $\vec{S}$ to $\vec{A}$ in fewer than $O(mnd)$ operations $\vec{S}$ must be structured. 
A common approach, used in many algorithms such as Blendenpik, is to make use of the fact that certain orthogonal transforms, such as the discrete cosine transform (DCT), can be applied to a length-$m$ vector in $O(m\log(m))$ time.
This yields a \emph{subsampled fast trigonometric transform} sketching matrix
\begin{equation}
\vec{S} = \vec{R}\vec{F}\vec{D}\vec{P},
\end{equation}
where $\vec{R}$ is a random restriction, $\vec{F}$ is an orthogonal trigonometric transform with a fast apply (e.g. DCT), $\vec{D}$ is diagonal matrix of random signs, and $\vec{P}$ is a random permutation matrix.
The sketch $\vec{S}\vec{A}$ can be computed in $O(mn\log(m))$ operations.

To satisfy theoretical guarantees \cref{thm:main}, subsampled fast trigonometric sketches require a sketch dimension $d = O(n\log(n))$, although in practice $d = O(n)$ seems to suffices in many situations \cite{martinsson_tropp_20}.
As such, the asymptotic runtime of \cref{alg:main} is reduced to $\tilde{O}(mn + n^3)$, substantially less than the $O(mn^2)$ operations required by classical direct solvers and Gaussian sketches.

Some downsides of subsampled fast trigonometric sketches are that they do not take advantage of possibly sparsity in $\vec{A}$ and that they have a $\log(m)$ dependence arising from the recursive structure of fast trigonometric transform algorithms.
This dependence is inherent to the approach, and cannot be reduced even if the user is willing to forgo theoretical guarantees.
The recursive nature of the algorithms also complicates efficient parallel implementation.

\subsection{Computational efficiency}
\label{sec:which_sketch:efficiency}

\begin{figure}[ht]
    \centering
    \includegraphics[scale=.7]{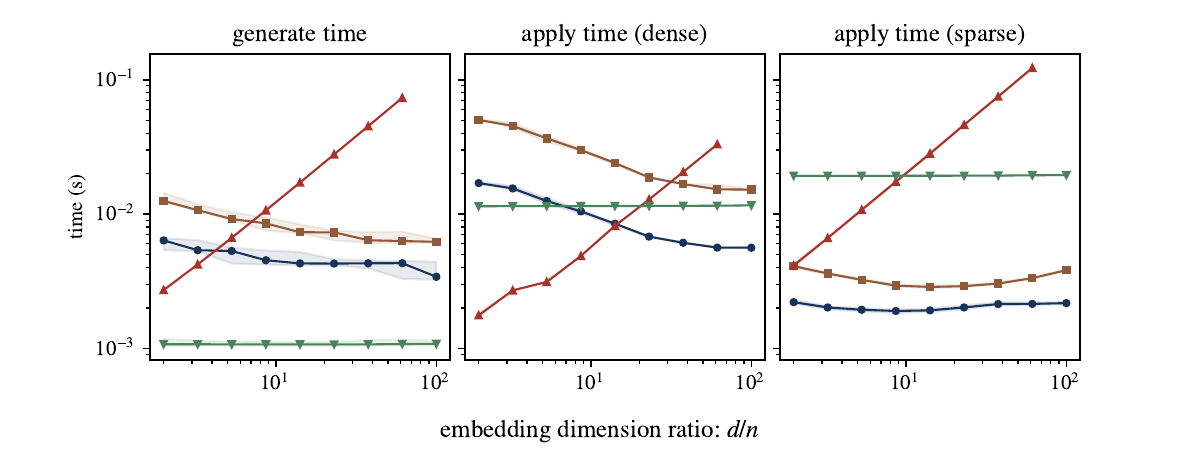}
    \caption{
    Time to generate (single GPU) and apply sketching matrices.
    Gaussian ({\protect\raisebox{.25pt}{\protect\includegraphics[scale=0.85]{imgs/legend/l2.pdf}}}), subsampled trig ({\protect\raisebox{.25pt}{\protect\includegraphics[scale=0.85]{imgs/legend/l3.pdf}}}), and sparse sign sketch with $\zeta = 8$ ({\protect\raisebox{.25pt}{\protect\includegraphics[scale=0.85]{imgs/legend/l1.pdf}}}) and $\zeta = 24$ ({\protect\raisebox{.25pt}{\protect\includegraphics[scale=0.85]{imgs/legend/l0.pdf}}}).
    For large embedding dimension, the Gaussian sketch required too much memory to generate, and is therefore not reported.
    }
    \label{fig:sketch_times}
\end{figure}

In the left panel of  \cref{fig:sketch_times}, we show the time required to generate the sketches described in \cref{sec:other_sketches}.
The time to generate a Gaussian sketch increases with the embedding dimension, as substantially more memory must be allocated and more random numbers sampled.\footnote{Since we only require the action of $\vec{S}$ on $\vec{A}$ and $\vec{b}$, the memory costs can be bypassed by generating and apply $\vec{S}$ at the same time, discarding the entries of $\vec{S}$ after they have been used.
However, this requires a more careful implementation. 
The use of low precision \cite{georgiou_boutsikas_drineas_anzt_23} can be used to reduce memory costs and build/apply time, but it will not change the scaling behavior.
}
On the other hand, the subsampled DCT is extremely fast to generate, and the generate time remains constant as the embedding dimension increases.

In the right two panels of \cref{fig:sketch_times}, we compare the time required  apply the sketches to a dense matrix $\vec{A}$ and to a sparse matrix $\vec{A}$.
In both cases, the cost of applying Gaussian sketch increases with the embedding dimension while the apply time for the subsampled DCT remains constant.
However, while the Gaussian sketch is substantially faster to apply when $\vec{A}$ is sparse, the subsampled DCT cannot take advantage of sparsity, and in fact is slightly slower than the dense case, due to having to allocate $mn$ memory.\footnote{Our implementation applies the fast DCT, and then subsamples. It may be possible to use a non-uniform DCT to avoid the large memory cost. However, implementations of the non-uniform DCT are less easily available and less optimized.}

\subsection{Distortion}
\label{sec:which_sketch:distortion}

\begin{figure}[ht]
    \centering
    \includegraphics[scale=.7]{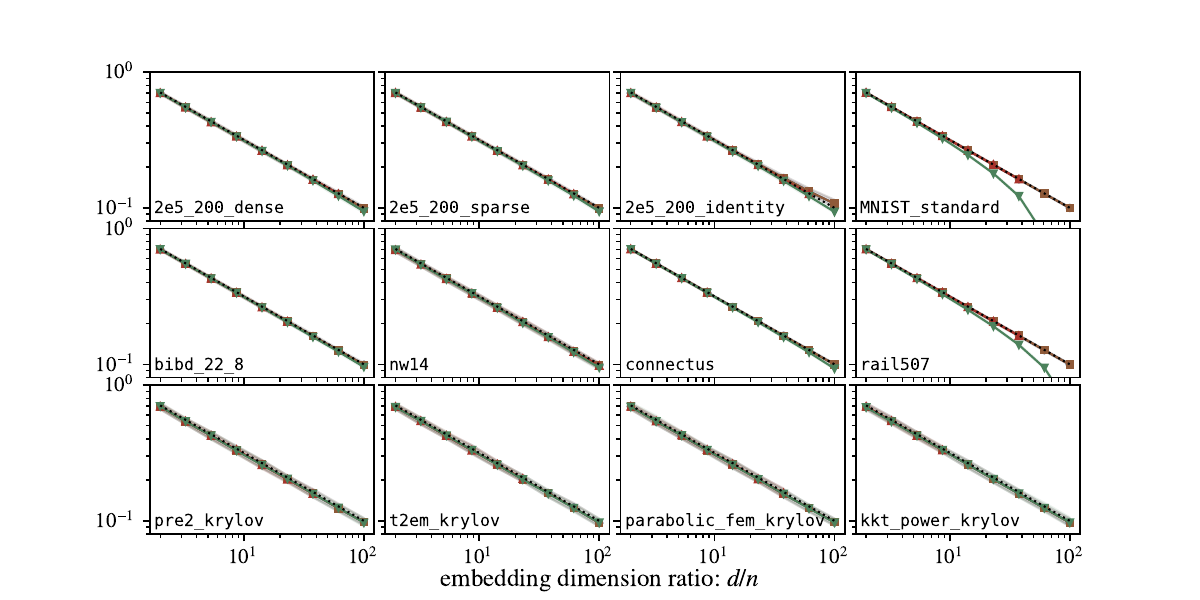}
    \caption{Distortion as a function of relative embedding dimension $d/n$ for different sketching distributions: 
    Gaussian ({\protect\raisebox{.25pt}{\protect\includegraphics[scale=0.85]{imgs/legend/l2.pdf}}}), subsampled trig ({\protect\raisebox{.25pt}{\protect\includegraphics[scale=0.85]{imgs/legend/l3.pdf}}}), and sparse sign sketch with $\zeta = 8$ ({\protect\raisebox{.25pt}{\protect\includegraphics[scale=0.85]{imgs/legend/l1.pdf}}}) and $\zeta = 24$ ({\protect\raisebox{.25pt}{\protect\includegraphics[scale=0.85]{imgs/legend/l0.pdf}}}).
    All sketches behave remarkably similarly to the asymptotic theory for Gaussian sketches which have distortion $\sqrt{n/d}$ ({\protect\raisebox{.25pt}{\protect\includegraphics[scale=0.85]{imgs/legend/dotted.pdf}}}).
    }
    \label{fig:sketch_distortion}
\end{figure}

In this section we repeat the experiment from \cref{sec:ed_sparsity_distortion} to understand the impact of embedding dimension on distortion for the sketching distributions described in \cref{sec:other_sketches}.
The results are shown in \cref{fig:sketch_distortion}.
For the most part, we observe that the sketches all behave similarly. 
The main exception is that the fast subsampled trigonometric sketches perform better than other when when $d\approx m$ since they have orthogonal rows.
However, the regime $d\approx m$ is not particularly useful within the sketch-and-precondition paradigm, as the cost of building the preconditioner would be comparable to the cost of a direct method.

\subsubsection{Full spectrum}

\begin{figure}[ht]
    \centering
    \includegraphics[scale=0.7]{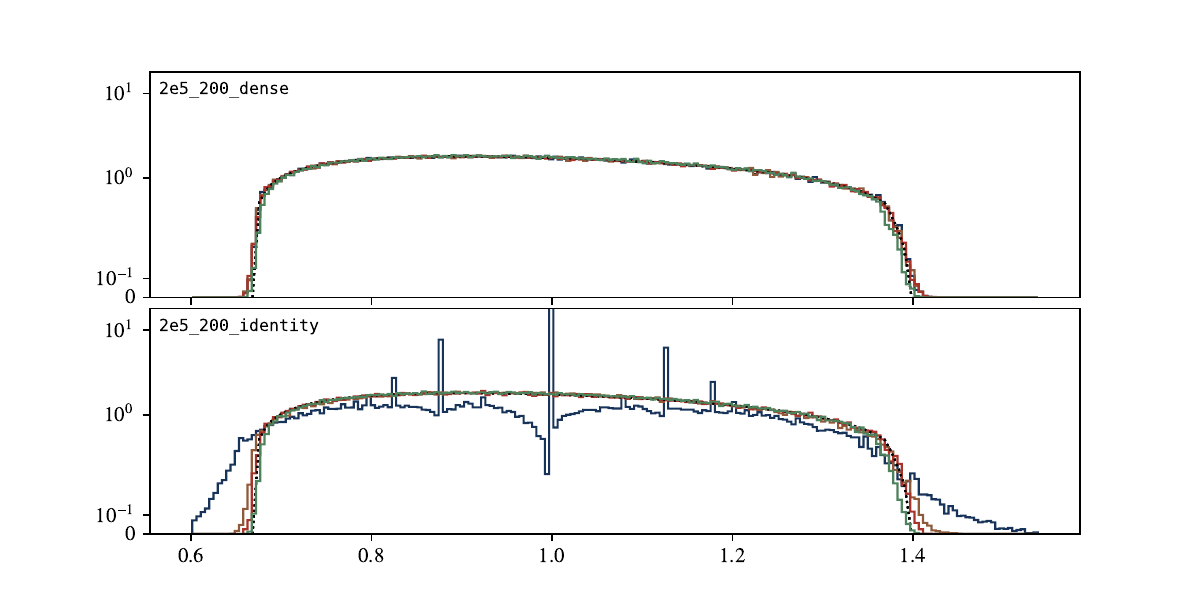}
    \caption{Spectrum of sketched subspace $\vec{S}\vec{U}$ on two test problems for various sketching distributions:
    Gaussian ({\protect\raisebox{.25pt}{\protect\includegraphics[scale=0.85]{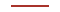}}}), subsampled trig ({\protect\raisebox{.25pt}{\protect\includegraphics[scale=0.85]{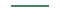}}}), and sparse sign sketch with $\zeta = 8$ ({\protect\raisebox{.25pt}{\protect\includegraphics[scale=0.85]{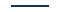}}}) and $\zeta = 24$ ({\protect\raisebox{.25pt}{\protect\includegraphics[scale=0.85]{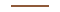}}}).
    Each histogram is constructed using 500 independent sketches.
    }
    \label{fig:spec_example}
\end{figure}

In order to probe the behavior further, in \cref{fig:spec_example} we plot the distribution of the squared singular values of $\vec{S}\vec{U}$, where $\vec{U}$ is any orthonormal basis for $\operatorname{range}(\vec{A})$.
For reference, we also plot the Marchenko--Pastur distribution with aspect ratio $d/n$.
This is the limiting behavior of a Gaussian sketch with aspect ratio $d/n$ held constant.

On a simple problem \texttt{2e5\_200\_dense}, whose left singular vectors are ``nice'', the sparsity parameter has little impact on the spectral properties of the sketch matrix $\vec{S}\vec{U}$, and in all cases, the spectrum is similar to what is predicted by Gaussian theory, with some deviation at the edges due to finite $n$ effects.
On the other hand, on the much harder \texttt{2e5\_200\_identity} example, we see the spectrum deviates greatly from the Gaussian limiting behavior when the embedding dimension is large and the sparsity is low. 
As the sparsity increases, the distribution much more well-behaved, but the edge behavior is still less defined.

\subsection{So, which sketch should I use?}

The experiments in \cref{sec:which_sketch:distortion,sec:which_sketch:efficiency} suggest that Gaussian, fast subsampled trigonometric, and sparse sign sketches all produces good subspace embeddings, and that subsampled trig and sparse sign sketches scale well with the embedding dimension.

While sparse sign sketches have an additional sparsity parameter, the experiments in \cref{fig:ed_sparsity_distortion} suggest that even a very aggressive parameter $\zeta = 8$ works well in many settings.
This aligns with past observations from the literature \cite{tropp_yurtsever_udell_cevher_19,dong_martinsson_23,epperly_23which}.
Critically, sparse sketches are efficient to apply and can take advantage of cases when $\vec{A}$ is itself sparse. 
On systems where $O(mn)$ memory is not available, this may be critical.
Finally, on distributed systems, the benefits of sparse sign sketches are even more pronounced.
In particular, the columns of a sparse sign sketch are independent, and hence the sketch generation and apply are embarrassingly parallel. 

Therefore, in line with \cite{epperly_23which,dong_martinsson_23,muarray_etal_23} \emph{we suggest sparse sign sketches as the default sketching distribution for most problems}.

\section{Krylov subspace methods}
\label{sec:KSMs}

It is easy to verify that the solution to \cref{eqn:lstsq_prec} is also the solution to the \emph{preconditioned normal equations}
\begin{equation}\label{eqn:normal}
\vec{Z}\vec{y} = \vec{f}
,\quad \vec{Z} = (\vec{A}\vec{M})^\T(\vec{A}\vec{M})
,\quad \vec{f} = (\vec{A}\vec{M})^\T\vec{b}
,\quad \vec{x} = \vec{M}\vec{y}.
\end{equation}
Given an initial guess $\vec{x}_0$, KSMs for \cref{eqn:lstsq_prec} produce an approximation $\vec{x}_t$ from the Krylov subspace
\begin{equation}\label{eqn:krylov_subspace}
\vec{x}_0 + \vec{M}\mathcal{K}_t(\vec{Z},\vec{r})
:=
\vec{x}_0 + \vec{M}\operatorname{span}\big\{ \vec{r}, \vec{Z}\vec{r}, \ldots, 
\vec{Z}^{t-1}\vec{r}\big\}
,\qquad \vec{r} = \vec{f} - \vec{Z}\vec{M}^{-1}\vec{x}_0.
\end{equation}
This space can be computed efficiently in an iterative manner. 
In particular, at each iteration, ones perform a matrix-vector product with $\vec{Z}$, which can be performed by sequential products with $\vec{M}$, $\vec{A}$, $\vec{A}^\T$, and $\vec{M}^\T$.

\subsection{Error analysis}
By definition, any vector $\vec{x}_t\in \vec{x}_0 + \vec{M} \mathcal{K}_t(\vec{Z},\vec{r})$ can be written as $\vec{x}_t = \vec{x}_0 + \vec{M}p_t(\vec{Z})\vec{r}$, where $p_t(x)$ is a polynomial of degree less than $t$. 
The true solution $\vec{x}_\star$ is expressed as $\vec{x}_\star = \vec{M} \vec{Z}^{-1}\vec{f} = \vec{x}_0 + \vec{M}\vec{Z}^{-1}\vec{r}$.
Hence, basic manipulations reveal that
\begin{align}
    \|\vec{A}(\vec{x}_\star - \vec{x}_t )\|
    &= \| \vec{A}\vec{M}(\vec{Z}^{-1}-p_t(\vec{Z}))\vec{r}\|
    \\&= \| \vec{Z}^{1/2}(\vec{Z}^{-1} - p_t(\vec{Z}))\vec{r}\|
    \\&= \| (\vec{I} - \vec{Z}p_t(\vec{Z})) \vec{Z}^{-1/2}  \vec{r} \|
    \\&\leq \| \vec{I} - \vec{Z}p_t(\vec{Z}) \| \| \vec{Z}^{-1/2} \vec{r} \|.
\end{align}
Now, observe that 
\begin{equation}    
    \| \vec{Z}^{-1/2} \vec{r} \|
    = \| \vec{Z}^{1/2} (\vec{Z}^{-1}\vec{f} - \vec{M}^{-1}\vec{x}_0) \|
    = \| \vec{A}\vec{M} (\vec{M}^{-1}\vec{x}_\star - \vec{M}^{-1}\vec{x}_0) \|
    = \| \vec{A}(\vec{x}_\star - \vec{x}_0) \|.
\end{equation}
In addition, since the spectral norm of a symmetric matrix is the largest absolute eigenvalue,
\begin{equation}
    \| \vec{I} - \vec{Z}p_t(\vec{Z})\|
    = \max_{x\in \operatorname{spec}(\vec{Z})} | 1-xp_t(x)|
    \leq \max_{x\in[\smin(\vec{Z}),\smax(\vec{Z})]} | 1-xp_t(x)|.
\end{equation}
In total, we therefore find that 
\begin{equation}
\label{eqn:KSM_error}
    \|\vec{A}(\vec{x}_\star - \vec{x}_t )\|
    \leq \bigg(\max_{x\in[\smin(\vec{Z}),\smax(\vec{Z})]} | 1-xp_t(x)| \bigg)  \| \vec{A}(\vec{x}_\star - \vec{x}_0) \|.
\end{equation}

There is a clear correspondence between different KSMs and different choices of $p_t(x)$.

\subsection{CG/LSQR}

The well-known conjugate gradient algorithm (which is algebraically equivalent to LSQR) applied to \cref{eqn:normal} produces iterates 
\begin{equation}\label{eqn:CG_def}
\vec{x}_t := \operatornamewithlimits{argmin}_{\vec{x}\in \vec{x}_0 + \vec{M}\mathcal{K}_t(\vec{Z},\vec{r})} \| \vec{A} (\vec{x}_\star-\vec{x}) \|
% = \min_{\deg(p)<t} \| (\vec{I} - \vec{Z}p_t(\vec{Z})) \vec{Z}^{-1/2}  \vec{r} \|
.
\end{equation}
By \cref{eqn:KSM_error}, we see that the CG iterates satisfy
\begin{equation}
    \|\vec{A}(\vec{x}_\star - \vec{x}_t )\|
\leq \min_{\deg(p)<t} \bigg(\max_{x\in[\smin(\vec{Z}),\smax(\vec{Z})]} | 1-xp(x)| \bigg)  \| \vec{A}(\vec{x}_\star - \vec{x}_0) \|.
\end{equation}
The polynomial minimization problem has a closed form solution in terms of Chebyshev polynomials, which gives the bound \cref{thm:CG_bound}.

Note that \cref{thm:CG_bound} holds for any $\vec{A}$, $\vec{b}$, $\vec{M}$, and $\vec{x}_0$; i.e., we have not made any use of the fact that the preconditioner $\vec{M}$ is obtained from sketching. 
At least in the case of a Gaussian sketch, the squared singular values of $\vec{A}\vec{M}$ have an inverse Wishart distribution.
This suggests that when $m$ and $n$ are large, the convergence of LSQR will  nearly deterministic \cite{deift_menon_olver_trogdon_14,deift_trogdon_20}, even in finite precision arithmetic 
\cite{chen_trogdon_24}.

\subsection{Gradient methods}
\label{sec:gd}

In place of LSQR, it is sometimes suggested to use a gradient-based KSM, such as gradient descent or Polyak's heavy ball momentum (\cref{alg:HBM}).
Standard analysis guarantees that, with the optimally chosen step size and momentum parameters $\alpha$ and $\beta$, that \cref{alg:HBM} produces iterates satisfying a guarantee very similar to \cref{thm:CG_bound}.
In particular, if 
\begin{equation}\label{eqn:hbm_params}
    \alpha = (1-\hat{\eta}^2)^2
    ,\qquad
    \beta = \hat{\eta}^2
    ,\qquad
    \cond(\vec{A}\vec{M}) \leq \frac{1+\hat{\eta}}{1-\hat{\eta}},
\end{equation}
then \cref{alg:HBM} converges at a rate $\sqrt{\beta}$:
\begin{equation}
    \lim_{t\to\infty}\left( \frac{\| \vec{A}(\vec{x}_\star - \vec{x}_t) \|}{ \| \vec{A}(\vec{x}_\star - \vec{x}_0) \|} \right)^{1/t}
    = \sqrt{\beta}.
\end{equation}
In the special case that $\hat{\eta}$ is such that $\cond(\vec{A}\vec{M}) = (1+\hat{\eta})/(1-\hat{\eta})$, then $\hat{\eta} = (\cond(\vec{A}\vec{M})-1)/(\cond(\vec{A}\vec{M})+1)$; i.e., the rate of convergence of \cref{alg:HBM} matches that of \cref{alg:LSQR}.
An implementation of heavy ball momentum is given in \cref{alg:HBM}.

\begin{algorithm}[ht!]
\caption{Gradient descent with momentum}
\label{alg:HBM}
	\begin{algorithmic}[1]
		\Require Matrix $\vec{A}\in\mathbb{R}^{m\times n}$, preconditioner $\vec{M}\in\mathbb{R}^{n\times n}$, vector $\vec{b}\in\mathbb{R}^m$, initial guess $\vec{x}_0\in\mathbb{R}^n$, accuracy target $\varepsilon$, step size $\alpha$, momentum $\beta$
        \State $\vec{x}_{-1} = \vec{x}_0$, $\vec{v} = \vec{M}\vec{M}^\T\vec{A}^\T\vec{b}$
        \For{$t=1,2,\ldots, $}
        \State $\vec{g}_t = \vec{v} - \vec{M}\vec{M}^\T \vec{A}^\T \vec{A}\vec{x}_{t-1}$
        \State $\vec{x}_t = \vec{x}_{t-1} + \alpha \vec{g}_t + \beta (\vec{x}_{t-1} - \vec{x}_{t-2})$
        \State Test for convergence and break when conditions are met
        \EndFor
	\Ensure Approximate solution $\widehat{\vec{x}} = \vec{x}_{t}$ to \cref{eqn:lstsq}
	\end{algorithmic}
\end{algorithm}

It is a standard exercise to show that \cref{eqn:hbm_params} gives the optimal choice of $\alpha$ and $\beta$, and other choices lead to either slower convergence or even divergence; see e.g. \cite{ozaslan_pilanci_arikan_19,pedregosa_21}.
Moreover, \cref{alg:HBM} produces iterates from \cref{eqn:krylov_subspace}, and hence the optimality of LSQR means that such methods cannot outperform LSQR in terms of iterations, even if the user is able to choose $\alpha$ and $\beta$ optimally. 
Therefore, in most situations, \emph{CG/LSQR should be preferred to HBM}.

The main setting in which HBM may be preferred is when matrix-vector products are comparatively cheap (e.g. because communication costs are high) \cite{meng_saunders_mahoney_14}.
Indeed, HBM has slightly lower arithmetic costs than LSQR, and more importantly, requires only one synchronization per iteration.
However, it is not hard to modify LSQR to reduce the synchronizations to one per iteration; see \cref{sec:LSQR_comm}.

\subsubsection{Numerical example}

\begin{figure}[ht]
    \centering
    \includegraphics[scale=.7]{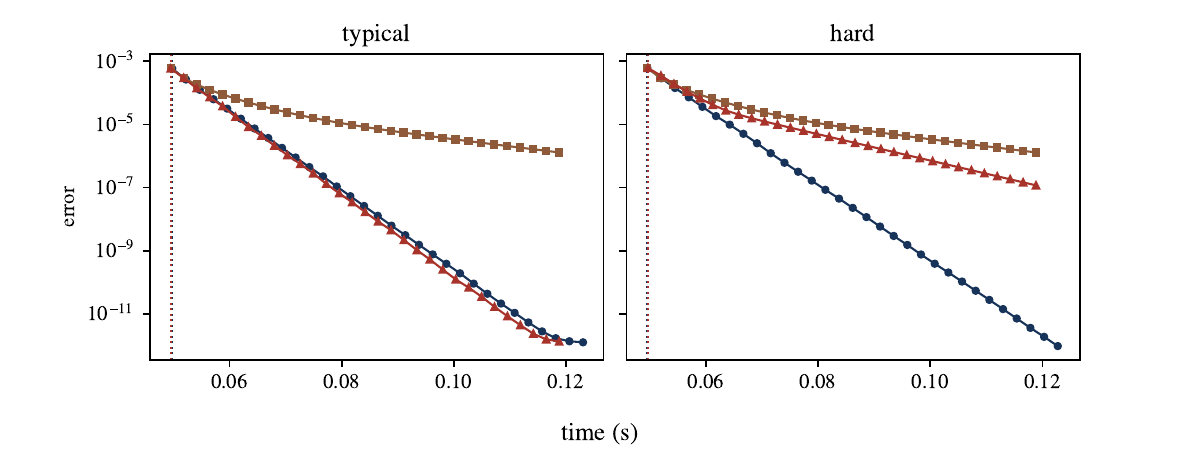}
    \caption{
    Error as a function of runtime for various iterative methods:
    LSQR ({\protect\raisebox{.25pt}{\protect\includegraphics[scale=0.85]{imgs/legend/l0.pdf}}}), 
    Gradient Descent
    ({\protect\raisebox{.25pt} {\protect\includegraphics[scale=0.85]{imgs/legend/l1.pdf}}}), and
    Heavy ball momentum ({\protect\raisebox{.25pt}{\protect\includegraphics[scale=0.85]{imgs/legend/l2.pdf}}}).
    }
    \label{fig:hbm_params}
\end{figure}

In \cref{fig:hbm_params} we show the convergence of LSQR and gradient descent (\cref{alg:HBM} with $\alpha= (1-\hat{\eta}^2)^2/(1+\hat{\eta}^2)$ and $\beta=0$) and heavy ball momentum with parameters \cref{eqn:hbm_params}.
Here we set $\hat{\eta} = \sqrt{n/d}$, based on the estimate for the distortion. 

As seen in the left panel, on typical problems where $\hat{\eta} = \sqrt{n/d}$ is a good estimate of the distortion (see \cref{fig:ed_sparsity_distortion}), HBM has a per-iteration convergence extremely similar to the optimal LSQR.
On this example, this is made up for by the slightly lower per-iteration cost.
On the other hand, on hard problems where $\hat{\eta} = \sqrt{n/d}$ is not a good estimate of the distortion, then HBM may converge more slowly (or even diverge). 
In both cases, gradient descent converges substantially slower.

To make HBM more robust, one can use an estimate $\hat{\eta} = c \sqrt{n/d}$ for some $c>1$ (in \cite{epperly_meier_nakatsukasa_24} it is suggested to set $c = \sqrt{1.1}$). 
However, this results in a slower rate of convergence.
On the other hand, LSQR does not require any sort of hyperparameter selection.

\subsection{Iterative sketching}
\label{eqn:iterative_sketching}

In \cite{pilanci_wainwright_16} and algorithm called 
\emph{iterative Hessian sketch} is introduced.
When applied to the least squares problem \cref{eqn:lstsq}, iterative Hessian sketch produces iterates
\begin{equation}
    \vec{x}_{t} = \argmin_{\vec{x}} \frac{1}{2d}\| \vec{S}_t \vec{A}(\vec{x} - \vec{x}_{t-1})\|^2
     + (\vec{x} - \vec{x}_{t-1})^\T\vec{A}^\T(\vec{b} - \vec{A}\vec{x}_{t-1}),
\end{equation}
where a fresh sketch $\vec{S}_i$ is drawn at each iteration. 
As with sketch-and-precondition, this algorithm converges in $t = O(1)\log(1/\varepsilon)$ iterations.
However, it requires generating and apply a fresh sketch every iteration, and solving a $d\times n$ dense linear algebra problem.
As a result, the total runtime is worse than sketch-and-precondition.

Subsequent work studied ``short-circuiting'' the iterative sketching approach by using a single sketch $\vec{S}$ in each iteration \cite{wang_lee_mahdavi_kolar_srebro_17,ozaslan_pilanci_arikan_19,lacotte_pilanci_21}.
Let $\vec{M} = \vec{R}^{-1}$ where $\vec{Q}\vec{R} = \Call{QR}{\vec{S}\vec{A}}$.
Then
\begin{align}
    \vec{x}_{t} 
    &= \argmin_{\vec{x}} \frac{1}{2d}\| \vec{S}\vec{A} (\vec{x} - \vec{x}_{t-1})\|^2
     + (\vec{x} - \vec{x}_{t-1})^\T\vec{A}^\T(\vec{b} - \vec{A}\vec{x}_{t-1})
    \\&=\argmin_{\vec{x}} \underbrace{\frac{1}{2d}\| \vec{M}^{-1} (\vec{x} - \vec{x}_{t-1})\|^2
     + (\vec{x} - \vec{x}_{t-1})^\T\vec{A}^\T(\vec{b} - \vec{A}\vec{x}_{t-1})}_{f(\vec{x})}
\end{align}
A direct computation reveals that the gradient of $f(\vec{x})$ is
\begin{equation}
    \nabla f(\vec{x})
    = \frac{1}{d}\vec{M}^{-\T}\vec{M}^{-1} (\vec{x} - \vec{x}_{t-1})
    + \vec{A}^\T(\vec{b} - \vec{A}\vec{x}_{t-1}).
\end{equation}
Therefore, we see that 
\begin{equation}
    \vec{x}_{t} = \vec{x}_{t-1} - \frac{1}{d}\vec{M} \vec{M}^\T \vec{A}^\T(\vec{b} - \vec{A}\vec{x}_{t-1}).
\end{equation}
In other words, this variant of iterative Hessian sketch is nothing more than standard gradient descent within the sketch-and-precondition framework. 
If momentum is used, as suggested in \cite{ozaslan_pilanci_arikan_19}, then one obtains sketch-and-precondition with heavy ball momentum (\cref{alg:HBM}).

As we discuss in \cref{sec:gd}, gradient-based methods, which require selection of certain hyperparameters, are unlikely to be a better choice than LSQR in most settings.

\subsection{Reducing communication in LSQR}
\label{sec:LSQR_comm}

Above, we note that gradient methods require fewer synchronizations per iteration than LSQR. 
However, with simple modifications, the LSQR iterate can be computed using a comparable number of synchronizations.
This is more well-studied for CG for square systems, where communication has an even larger impact; see e.g. \cite{meurant_87,chronopoulos_gear_89,carson_15}.

\subsubsection{Hiding communication}

The first approach is to simply restructure the LSQR iteration so that the communication heavy operations can be overlapped. 
Recall \cref{alg:LSQR} uses updates:
\par\addvspace{-.5em}\begin{minipage}{\linewidth}
\hrule\kern2pt
\begin{algorithmic}[1]
    \setcounter{ALG@line}{2}
    \State $\hat{\vec{u}}_{t+1} = \vec{A}\vec{M}\vec{v}_{t} - \alpha_{t} \vec{u}_{t}$
    \State $\vec{u}_{t+1} = \hat{\vec{u}}_{t+1} / \beta_{t+1}$,~ $\beta_{t+1} = \| \hat{\vec{u}}_{t+1}\|$ 
    \State $\hat{\vec{v}}_{t+1} = \vec{M}^\T\vec{A}^\T\vec{u}_{t+1} - \beta_{t+1}\vec{v}_{t}$
    \State $\vec{v}_{t+1} = \hat{\vec{v}}_{t+1} / \alpha_{t+1} $,~ $\alpha_{t=1} = \| \hat{\vec{v}}_{t+1}\|$.
\end{algorithmic}
\hrule\kern2pt
\end{minipage}

When parallelized as described in \cref{sec:our_implementation:parallel}, this requires two synchronizations between parallel resources. 
In particular, computing $\beta_{t+1} = \| \hat{\vec{u}}_{t+1}\|$ requires collecting the partial norms from each chunk of $\hat{\vec{u}}_{t+1}$. 
Then computing $\vec{A}^\T\vec{u}_{t+1}$ also requires a synchronization. 

This can be brought down to one synchronization by reorganize the iteration as:
\par\addvspace{-.5em}\begin{minipage}{\linewidth}
\hrule\kern2pt
\begin{algorithmic}[1]
    \setcounter{ALG@line}{2}
    \State $\hat{\vec{u}}_{t+1} = \vec{A}\vec{M}\vec{v}_{t} - \alpha_{t} \vec{u}_{t}$
    \State $\tilde{\vec{v}}_{t+1} = \vec{M}^\T\vec{A}^\T\hat{\vec{u}}_{t+1}$,~ $\vec{u}_{t+1} = \hat{\vec{u}}_{t+1} / \beta_{t+1}$,~ $\beta_{t+1} = \| \hat{\vec{u}}_{t-1}\|$ 
    \State $\hat{\vec{v}}_{t+1}  = \tilde{\vec{v}}_{t+1} / \beta_{t+1} - \beta_{t+1}\vec{v}_{t}$
    \State $\vec{v}_{t+1} = \hat{\vec{v}}_{t+1} / \alpha_{t+1} $,~ $\alpha_{t=1} = \| \hat{\vec{v}}_{t+1}\|$.
\end{algorithmic}  
\hrule\kern2pt
\end{minipage}

\subsubsection{Avoiding communication}
Let $\vec{q}_1, \ldots, \vec{q}_t$ be a basis for $\vec{M}\vec{K}_t(\vec{Z},\vec{r})$ and define $\vec{Q} = [\vec{q}_1, \ldots, \vec{q}_t]$.
Then, the CG iterate \cref{eqn:CG_def} has a closed-form expression
\begin{equation}\label{eqn:CG_closedform}
    \vec{x}_t = 
\vec{x}_0 + \vec{Q}_t (\vec{Q}_t^\T \vec{A}^\T\vec{A}\vec{Q}_t)^{-1} \vec{Q}_t^\T\vec{A}^\T \vec{r}.
\end{equation}
In particular, if $\vec{q}_1, \ldots, \vec{q}_t$ are some explcit polynomial basis (e.g. corresponding to the Chebyshev polynomials) that can be computed in an inner-product free way, then \cref{eqn:CG_closedform} allows the CG iterate to be computed with few additional synchronizations. 
This is related to so-called $s$-step KSMs \cite{chronopoulos_gear_89}.

Typically, the main issue with \cref{eqn:CG_closedform} is that the numerical behavior can be poor if $\vec{Q}$ is ill-conditioned. 
However, owing to the fact that we have a very good idea of key properties of the spectrum of $\vec{Z}$, obtaining a well-conditioned basis seems within reach.

\end{document}